\documentstyle[aps,preprint,epsf]{revtex}

\def\e{\mbox{e}}

\begin{document}
\draft
\preprint{UW/PT-96-05}
\title{Reheating and thermalization in a simple scalar model}
\author{D.~T.~Son}
\address{Department of Physics FM--15, University of Washington,
Seattle, WA 98105, USA}
\date{April 1996}
\maketitle
\begin{abstract}
We consider a simple model for the Universe reheating, which consists
of a single self--interacting scalar field in Minkowskian space--time.
Making use of the existence of an additional small parameter
proportional to the amplitude of the initial spatially homogeneous
field oscillations, we show that the behavior of the field can be
found reliably.  We describe the evolution of the system from the
homogeneous oscillations to the moment when thermalization is
completed.  We compare our results with the Hartree--Fock
approximation and argue that some properties found for this model may
be the common features of realistic theories.

\end{abstract}


\section{Introduction}

The problem of decay of homogeneous oscillations of the inflaton field
after inflation is one of the most important issues of inflationary
cosmology.  This process is responsible for the reheating of the
Universe, in particular, the reheating temperature depends on the
details of the decay of homogeneous oscillations.  Recently, this
problem has attracted considerable interest, and new investigations
have revealed a rather complicated dynamics of the inflaton field.  In
the old approach, which is called ``elementary theory'' \cite{old},
the decay of the spatially homogeneous oscillations of the inflaton
field is described by a friction term in the equation of motion, which
has a coefficient proportional to the decay rate of the inflaton
quanta.  This picture leads to a slow, exponential decay of the
amplitude of oscillations.  It is not until recently that it has been
discovered that in most important cases this picture is not correct.
The reason for that is usually related to the existence of the
resonance bands in the spectrum of the scalars interacting with the
inflaton field, which contains modes that are unstable on the
background of homogeneous oscillations of inflaton field.  The
importance of parametric resonance was first realized by Kofman, Linde
and Starobinsky \cite{KLS}.  The modes affected by this phenomenon may
be of the inflaton field itself or belongs to any scalar coupled to
the inflaton.

As the importance of the parametric resonance has been understood, a
new scenario has be proposed \cite{KLS} (see also \cite{Shtanov}).
According to the latter, the reheating process occurs in two steps.
At the first step, dubbed ``preheating'', the amplitudes of the
quantum modes inside the resonance bands increase exponentially,
giving rise at some time moment to the creation of a large number of
quanta of inflaton or other scalar fields.  Due to the exponential
character of the growth of the resonance modes, the creation of the
particles by parametric resonance is an explosion--type process, where
most particles are emitted at the very last moments of the preheating
epoch.  After this explosion, most energy is carried by the created
particles, and the old theory of reheating can be applied to the
thermalization of the products of explosion and to the decay of the
remnant of the homogeneous oscillations.

Despite the fact that there is a growing number of works aimed at
extracting physical consequences from the scenario advocated in
Ref.\cite{KLS} (see Refs.\cite{KLS2,Tkachev,Kolb}), the latter
deserves more detail theoretical elaboration in order to take into
account important physical effects that may change many details of the
evolution of the Universe after the explosion.  Let us mention only
two of these effects.  The first is the phase coherence between the
modes created by parametric resonance: the wave functions of particles
created in this way have almost the same phase, which is correlated
with the phase of the homogeneous oscillations.  This phase
correlation, presumably, may have certain consequences, in the same
manner as the coherent nature of the homogeneous oscillations lead to
the parametric resonance.  The second important effect is the
scattering between the particles, which cannot be neglected even in
theories with small coupling.  In fact, according to the standard
picture, the particles created by preheating have very large
occupation numbers, typically of the order of the inversed coupling.
These large occupation numbers lead to strong Bose enhancement of the
scattering processes, and the system is effectively strongly coupled.
This implies, in particular, that after the explosion, the relaxation
time is of the same order as the period of oscillation, and the effect
of particle collision cannot be neglected.  Existing calculations
\cite{KLS,Boyan} mostly rely on the Hartree--Fock approximations,
where the effect of particle collision is not taken into account.

Closely related to the decay of homogeneous oscillation is the
thermalization among products of parametric resonance particle
production.  It has been argued that since the particles produced by
parametric resonance has energy much smaller than the equilibrium
temperature, the thermalization takes a long time interval as the
particles have to go through may collisions in order to overcome the
wide separation of energy scales.  However, no theory of
thermalization has been developed, as well as no reliable estimation
of the thermalization time has been made so far.

That the Universe after the preheating explosion is effectively
strongly--coupled as mentioned above makes the study of the dynamics
of the Universe quite difficult, and it has been suggested that
numerical methods may be used to simulate processes occurring after
inflation \cite{Son}.  Numerical simulations of this type has been
carried out for the simplest $\phi^4$ theory in Ref.\cite{Khlebnikov},
but in order to understand the qualitative features in the evolution
of the Universe after inflation, as well as to check the validity of
the Hartree--Fock approximation, it is desirable to have a simple
model where this evolution can be followed in the whole time interval
from the beginning up to the thermalization in a reliable way.

In this paper we consider one of such models.  We neglect the
expansion of the Universe, and consider the theory of one
self--coupled scalar filed without symmetry breaking.  The initial
amplitude of the homogeneous oscillations is taken to be small enough
for the latter to be almost sinusoidal.  More precisely, the amplitude
is supposed to be $A_0m/\sqrt{\lambda}$, where $m$ is the boson mass
and $\lambda$ is the coupling, and $A_0$ is some small dimensionless
parameter, $A_0\ll1$.  Despite the fact that the effect of parametric
resonance is suppressed by a power of $A_0$, we will show that the
analogue of the elementary theory of reheating is totally inapplicable
when $A_0\gg\sqrt{\lambda}$.  In particular, this theory is invalid in
the regime that we consider throughout the paper, where $A_0$ is
small, but fixed, and $\lambda\to0$.  By considering the regime where
the elementary theory breaks down, we let our model possess the main
feature of the theories where parametric resonance is important.
That the evolution of the Universe can be found reliably in our regime
is the consequence of the existence of an additional small parameter
$A_0$ in our model.

Let us briefly outline our results.  At first, when the field makes
homogeneous oscillations, the modes inside the resonance bands are
enhanced and their amplitudes increase exponentially with time.  The
modes that are enhanced most are those around the center of the
resonance band near $2m$.  At some time moment, when the energy
carried by the resonance modes is still small, the back-reaction of
these modes to the homogeneous background and their self--interaction
become important and the exponential growth of the modes terminates.
The dynamics of the field becomes rather complicated, but during some
time interval it can be treated by the Hartree--Fock approximation.
We see that the coherent nature of the modes created by parametric
resonance leads to a period of time where some portion of the energy
is transfered back and forth between the homogeneous oscillations and
the resonance modes.  Therefore, the decay rate is reduced by the
coherence effect.  After some time interval the coherence is
destroyed, and the system goes to another regime where created modes
are uncorrelated.  In this regime, there is still coherent enhancement
of modes by parametric resonance, but now the width of the resonance
band is much smaller than that of the particle spectrum.  Moreover,
the resonance band moves towards the region of smaller energies, so
each mode with a given momentum is enhanced during only a finite time
interval, after than they become uncorrelated and can be considered as
ordinary particles.  By this mechanism, the long--run effect of
parametric resonance is non--explosive production of particles with
energy approximately $2m$.  The rate of particle production slowly
changes with time, but not by many orders of magnitude.

After a period of particle emission, when most energy is still carried
by the homogeneous mode, the density of created particles is already
enough for the non--coherent scattering between the particles becomes
important and the Hartree--Fock approximation breaks down.  To
describe the system in this regime, we make use of the Boltzmann
kinetic equation.  We find that the production of particles via
parametric resonance occurs concurrently with their dissipation to
other modes by scattering, which leads to the excitation of modes with
larger and larger energies.  It is very interesting that the dynamics
of the scalar field is similar to that of turbulent systems, where
energy is transfered from long--distance to short--distance modes.  In
particular, the Kolmogorov power spectrum is formed after some time
interval among the particles with nonzero momenta.

During the whole period describe above, the amplitude of homogeneous
oscillations gradually decreases.  When the energy carried by the
homogeneous mode becomes much smaller than the total energy, the
thermalization epoch begins.  The thermalization can be divided into
two steps where the distribution function shows two different
self--similar behaviors, depending on whether the Bose condensation is
effective compared to condensate evaporation.  Let us only note that
the typical energy of particles increases continuously, while the
typical occupation number decreases.  The time required to achieve
thermal equilibrium is estimated to be $t\sim
A_0^{1/6}\lambda^{-7/4}m^{-1}$.

Therefore, the results of our paper show that the dynamics of the
scalar field after the explosion is much more complicated than
predicted by the Hartree--Fock approximation, and there is interplay
between various processes, including particle creation, scattering,
and Bose condensation, which must be accurately taken into account in
order to find out the actual behavior of the system.  Generally, we
find that the scattering slows down the decay of the homogeneous
oscillations.  Our investigation suggests that features discovered in
our model may be common in some realistic theories.

The paper is organized as follows.  In Sect.2 we describe the model,
and show that the elementary theory of reheating is not valid in a
particular regime in this model.  We consider the parametric
resonance, in particular the strongest resonance band that is located
at energies closed to $2m$.  We show that the linear regime, where
back-reaction of the created particle to the background can be
neglected, breaks down when the energy carried by produced particles
is still small.  Sect.3 is devoted to the Hartree--Fock approximation,
where particles are considered as moving on a self--consistent
background.  We derived the Hartree--Fock equation for the amplitude of
the wave functions of the particles, and show that the phase coherence
of the latter slow down the decay of the homogeneous oscillations.  We
find analytically the rate of particle creation in the long run.  In
Sect.4 we take into account the scattering among produced particles
and describe the evolution of the system in the regime where both
particle creation and scattering occurs at the same speed.  We show
that the part of our system that contains particles with nonzero
momenta behaves like a typical turbulent system.  In particular, we
find the Kolmogorov index of turbulence for our case.  In Sect.5 we
discuss the behavior of the system after the amplitude of homogeneous
oscillations become small is considered.  The thermalization is
described, and the time required for it is also estimated.  Finally,
Sect.6 contains concluding remarks.

\section{The model}

Let us consider the theory of one scalar field in flat Minkowskian
space--time,
\begin{equation}
  L={1\over2}(\partial_\mu\phi)^2-{1\over2}\phi^2-{\lambda\over4!}\phi^4
  \label{Lagrangian}
\end{equation}
where the coupling constant $\lambda$ is assumed to be small,
$\lambda\ll1$.  For convenience we take the mass of the boson to be 1.
We suppose that at the beginning the field $\phi$ makes spatially
homogeneous oscillations.  This can be done by adding a source term
$J\phi$, where $J$ is a constant at $t<0$, to the Lagrangian
(\ref{Lagrangian}) and turn off $J$ at $t=0$.  We assume that the
amplitude of the oscillation is much smaller than $1/\sqrt{\lambda}$,
so that it is almost sinusoidal.  To the leading approximation, the
time dependence $\phi$ is of the form
\begin{equation}
  \phi(t)={2\over\sqrt{\lambda}}A_0\cos t
  \label{cos}
\end{equation}
where $A_0\ll1$.  We will be interested in the process of energy
transfer from the homogeneous mode to the modes with nonzero momentum.
The regime that we are interested in is small, but fixed, $A_0$, and
$\lambda\to0$.  Let us first show that in this regime, the elementary
theory of reheating is not applicable.

\subsection{Elementary theory}

The homogeneous oscillation (\ref{cos}) can be considered as a
condensate of particles with zero momentum.  The particle density in
the condensate that corresponds to the oscillations (\ref{cos}) is
\begin{equation}
  n_0={2A_0^2\over\lambda}
  \label{n0}
\end{equation}
In the elementary theory of reheating, the decay of the homogeneous
mode (which will also be called ``condensate evaporation'') is viewed
as a perturbative process which usually occurs via the decay of
inflaton quanta in the condensate to other particles.  In our case
where the theory contains only one scalar, the $\phi$--quanta are
stable, but there are still processes that creates $\phi$--quanta with
nonzero momentum from those in condensate.  The first perturbative
process of this type is the $4\to2$ process, in which 4 $\phi$--bosons
at rest become to 2 particles with energy $E=2$ each.  The amplitude
of this $4\to2$ process is not vanishing and proportional to
$\lambda^2$.  The rate of the $4\to2$ can be obtained by multiplying
the squared amplitude to the fourth power of $n_0$ (since it involves
4 initial particles), and the elementary theory in our case predicts
\begin{equation}
  {dn_0\over dt}\sim\lambda^4n_0^4\sim A_0^8
  \label{dn0/dt}
\end{equation}
The time scale of the decay of the condensate is
\begin{equation}
  T\sim n_0\left({dn_0\over dt}\right)^{-1}\sim
  {1\over\lambda A_0^6}
  \label{naive_timescale}
\end{equation}
However this naive calculation is incorrect.  Though usually the
breakdown of the elementary theory is related to the phenomenon of
parametric resonance, it can be easily understood also in the
perturbative framework.  To do that, let us recall that in our
calculations we assume implicitly that the two final particles of the
$4\to2$ process are emitted as in vacuum.  However, the rate of this
process is enhanced if there are already particles with energy $E=2$.
In fact, provided the occupation number at the energy level $E=2$ is
$n$, then the $4\to2$ probability is enhanced by the Bose factor of
$(1+n)^2$.  There is also the time--reversal transition where two
particles with energy $E=2$ go back to 4 particles at rest, which has
a relative rate of $n^2$.  If one combines the two processes one finds
that the overall rate of transition from the condensate is
$(1+n)^2-n^2=1+2n$ times larger than predicted by the elementary
theory, which implies that the latter is correct only when the
occupation number $n$ is much smaller than 1.

Let us estimate the occupation number at the energy level $E=2$.  At
the beginning this occupation number is 0, but it increases due to the
radiation by $4\to2$ transitions.  The particles with energy $E=2$
would be accumulated in large number if they remain untouched at the
same energy level after they have been created, however in reality
these particles scatter on those in the condensate and have a finite
mean free time of order
\[
  t_0\sim{1\over\sigma n_0}\sim{1\over\lambda A_0^2}
\]
(where $\sigma$ is the typical cross section in particle collisions,
$\sigma\sim\lambda^2$).  Note that at the values of $A_0$ we are
considering, $A_0\ll1$, the mean free time $t_0$ is much smaller than
the time scale in eq.(\ref{naive_timescale}).  Each particle created
from the condensate lives at the energy level $E=2$ for a time
interval $t_0$, therefore the maximal density of particles with energy
$E=2$ that can be achieved is
\begin{equation}
  \rho(E=2)\sim{dn_0\over dt}\cdot t_0\sim{A_0^6\over\lambda}
  \label{density}
\end{equation}
Though this density is much smaller than $n_0$, the occupation number
would be infinite if all these particles have exactly the same energy.
However, due to the Heisenberg principle, the energy of particles has
an uncertainty of order $t_0^{-1}$ since they have a finite lifetime.
The particle distribution in the momentum space, so, can be imagined
as a spherical shell around $E=2$ with the thickness of $t_0^{-1}$.
The occupation number can be found by dividing the particle density
(\ref{density}) to the phase space that the particles occupy,
\[
  n\sim{\rho(E=2)\over t_0^{-1}}\sim{A_0^4\over\lambda^2}
\]
Recall that the condition for the elementary theory of reheating to be
valid is $n\ll1$, one finds that the latter can be applied only when
$A_0\ll\sqrt{\lambda}$, or, in other words, the amplitude of the
homogeneous oscillations must be much smaller than 1.  Since in the
regime considered in this paper ($A_0=\mbox{fixed}$, $\lambda\to0$)
this condition is violated, the elementary theory is not appropriate
for our purposes.

\subsection{Parametric resonance}

Since the elementary theory of reheating breaks down, let us turn to
the parametric resonance.  For further calculations, we need a
more accurate expression for the solution to the field equation than
eq.(\ref{cos}).  At small $A_0$, the following formula for the solution
can be derived,
\begin{equation}
  \phi_0(t)={1\over\sqrt{\lambda}}\left(2A_0\cos\omega t+
  {A_0^3\over24}\cos3\omega t+O(A_0^5)\right)
  \label{sol}
\end{equation}
where $\omega=1+{A_0^2\over4}+O(A_0^4)$ is the frequency of the
oscillations.  As noted above, the oscillations is almost a cosine:
the higher harmonics come with smaller coefficients than the leading
one.

If the system under consideration is classical, and the spatial
homogeneity is exact, the oscillations (\ref{sol}) would continue
infinitely long.  However, this is not true in the quantum case.  The
reason for the behavior of the quantum system to be different is that
the homogeneous oscillation (\ref{sol}) is unstable: some vacuum
quantum fluctuations become larger with time.  To see this instability
let us quantize the scalar field on the background (\ref{sol}).  For
this end we decompose the file operator into classical and quantum
parts, $\phi=\phi_0+\tilde{\phi}$.  The part of the Lagrangian that is
quadratic on $\tilde{\phi}$ has the form,
\begin{equation}
  L_2[\tilde{\phi}]={1\over2}(\partial_\mu\tilde{\phi})^2-
  {1\over2}\left(1+{\lambda\over2}\phi_0^2(t)\right)\tilde{\phi}^2
  \label{L2}
\end{equation}
The quantization is completely analogous to that of the free scalar
field on the vacuum background.  As the result, one represents
$\tilde{\phi}$ via the creation and annihilation operators
\begin{equation}
  \tilde{\phi}(x)=\int\!{d{\bf k}\over(2\pi)^32\omega_k}\,
  \left(a_{\bf k}f(k,t)+a^\dagger_{\bf k}f^*(k,t)\right)
  \label{quantiz}
\end{equation}
where $[a_{\bf k}, a^\dagger_{\bf k'}]=(2\pi)^32\omega_k\delta({\bf
k}-{\bf k}')$.  If there is no homogeneous oscillations, one has
$f(k,t)=\e^{-i\omega_kt}$, where $\omega_k=\sqrt{k^2+1}$, and
eq.(\ref{quantiz}) reduces to the usual formula for quantized scalar
field on the vacuum background.  On the background of $\phi_0$,
$f(k,t)$ is a solution to the mode equation
\begin{equation}
  \left(\partial_t^2+\omega_k^2+
  {\lambda\over2}\phi_0^2(t)\right)f(k,t)=0
  \label{mode_eq}
\end{equation}
To find the initial condition for eq.(\ref{mode_eq}), let us note that
at $t<0$ the source term $J\phi$ is present in the Lagrangian, so
$\phi_0$ is constant, and the Lagrangian (\ref{L2}) is just that of a
scalar field on the vacuum background.  Neglecting the small
correction to the mass, one has
\begin{equation}
  f(k,t)=\e^{-i\omega_kt}, \qquad t<0
  \label{fekt}
\end{equation}
We will call $f(k,t)$ the wave function of particles with momentum
${\bf k}$ (in fact it gives the time dependent of the wave function).
Eq.(\ref{mode_eq}) has the same form as the equation of motion of an
oscillator whose frequency is a periodic function of time,
$\omega^2(t)=\omega_k^2+{\lambda\over2}\phi_0^2(t)$.  Such systems are
well known in classical mechanics to exhibits the phenomenon of
parametric resonance, which means that there are values of $k$ where
any solution to $f(k,t)$ to eq.(\ref{fekt}) becomes arbitrarily large
in at least one of the limits $t\to-\infty$ or $t\to+\infty$
\cite{Landau}.  In the case when the dependence of the frequency on
time is weak, as in our problem ($A_0\ll1$), the solution to
eq.(\ref{mode_eq}) can be found analytically, and the parametric
resonance occurs when $\omega_k$ is close to 1, 2, 3, etc.\ (the
higher $\omega_k$, the weaker is the parametric resonance) .  To
demonstrate the technique which will be extensively explored in this
paper, let us discuss the first resonance near $\omega_k\approx1$.

\subsection{First resonance band}

Let us follow the technique of \cite{Landau} and find out the first
resonance band near $\omega_k\approx1$, where the parametric resonance
is strongest.  The result in this subsection will not be used in other
part of the papers, sine the resonance band lies in the region of
unphysical values of $\omega_k$.  However, the techniques used here to
find the first resonance band can be applied, with minor
modifications, to the second resonance and the Hartree--Fock
approximation, so they are worth detail consideration.

First, let us substitute eq.(\ref{sol}) to eq.(\ref{mode_eq}).  For
our purpose, only the leading term in eq.(\ref{sol}) is important.
One obtains,
\begin{equation}
  \left(\partial_t^2+\omega_k^2+A_0^2+
  {A_0^2\over2}\e^{-i\omega t}+{A_0^2\over2}\e^{i\omega t}
  \right)f(k,t)=0
  \label{1res}
\end{equation}
The idea is to find $f(k,t)$ in the form of a function with
oscillatory behavior with the same frequency $\omega$ as $\phi_0$, but
the amplitude of the oscillations is supposed to be a slowly varying
function of time.  So we look for the solution in the form
\begin{equation}
  f(k,t)=\alpha(t)\e^{-i\omega t}+\alpha^*(t)\e^{i\omega_k t}+\cdots
  \label{1res_sol}
\end{equation}
where $\alpha(t)$ varies in a time scale much larger than 1, and dots
stay for terms of higher harmonics.  Substituting eq.(\ref{1res_sol})
to eq.(\ref{1res}), and neglecting terms with second derivatives of
$a$ and $b$ with time, as well as those with higher harmonics, we
obtain
\begin{equation}
  \left(-2i\omega\dot{\alpha}+(\omega_k^2-\omega^2+A_0^2)\alpha+
  {A_0^2\over2}\alpha^*\right)\e^{-i\omega t}+
  \left(2i\omega\dot{\alpha}^*+(\omega_k^2-\omega^2+A_0^2)\alpha^*
  +{A_0^2\over2}\alpha\right)\e^{i\omega t}=0
  \label{ikt-ikt}
\end{equation}
Since $\alpha$ and $\alpha^*$ vary much slower than $\e^{-i\omega t}$
and $\e^{i\omega t}$, eq.(\ref{ikt-ikt}) can be satisfied only if both
the coefficients of $\e^{-i\omega t}$ and $\e^{i\omega t}$ vanish.
Thus, one obtains the following equations,
\begin{equation}
  -2i\omega\dot{\alpha}+(\omega_k^2-\omega^2+A_0^2)\alpha+
  {A_0^2\over2}\alpha^*=0
  \label{1_res}
\end{equation}
\[
  2i\omega\dot{\alpha}^*+(\omega_k^2-\omega^2+A_0^2)\alpha^*
  +{A_0^2\over2}\alpha=0
\]
The solution to this set of equation can be searched in the form
$\alpha(t)=\alpha_0\e^{s_kt}$, $\alpha^*(t)=\alpha^*_0\e^{s_kt}$.  One
finds
\begin{equation}
  s_k=\pm{1\over2\omega}\sqrt{\left({A_0^2\over2}\right)^2-
  (\omega_k^2-(\omega^2-A_0^2))^2}
  \label{sk1}
\end{equation}
Note that if one considers only the values of $\omega_k$ so that
$\omega_k-\omega\sim A_0^2$, then $s_k\sim A_0^2\ll1$, and our
assumption that $\alpha$ varies on a time scale much larger than 1 is
justified (in our case this time scale is $A_0^{-2}$).  Depending on
whether $|\omega_k^2-\omega^2+A_0^2|$ is larger or smaller than
$A_0^2/2$, $s_k$ is purely imaginary or real.  If $s_k$ is purely
imaginary, the solutions cannot be arbitrary large.  The parametric
resonance occurs only when $s_k$ is real, for example, the solution
corresponding to positive $s_k$ can be arbitrary large if $t$ is large
enough.  From eq.(\ref{sk1}) one finds that the parametric resonance
occurs when $\omega_k$ lies in a finite window,
\begin{equation}
  \omega^2-{3A_0^2\over2}<\omega_k^2<\omega^2-{A_0^2\over2}
  \label{range}
\end{equation}
However, it is easy to note that since $\omega^2=1+{A_0^2\over2}$, the
upper limit of the range (\ref{range}) coincides with 1.  So,
eq.(\ref{range}) corresponds to unphysical values of $\omega_k$,
$\omega_k<1$, and there is no parametric resonance at physical
$\omega_k$.

So far, our calculations are performed only to the first order on
$A_0^2$ and do not exclude the possibility that the upper limit of the
resonance band is differ from 1 by a value of $O(A_0^4)$, so there is
still a very small region of physical $\omega_k$ where there is
parametric resonance.  However, the coincidence of the upper limit of
the band and the boson mass can be shown to be hold exactly, to any
order of $A_0$.  In fact, at $\omega_k=1$, the mode equation
(\ref{mode_eq}) possesses a solution $f(0,t)=\dot{\phi}_0(t)$ that is
a periodic function of $t$ with the same period as $\phi_0$.  This
fact shows that at $k=0$ one has $s_k=0$ exactly.

\subsection{Second resonance band}

Let us turn to the second resonance at $\omega_k\approx2$.  Now one
must take into account the second term in the expansion (\ref{sol})
and the linearized equation for $f(k,t)$ is
\begin{equation}
  (\partial_t^2+\omega_k^2+A_0^2)f(k,t)+\left({A_0^2\over2}
  \e^{-2i\omega t}+{A_0^4\over48}\e^{-4i\omega t}+\mbox{h.c.}
  \right)f(k,t)=0
  \label{2nd_res}
\end{equation}
In analogy with the case of first resonance band, we will seek the
solution to eq.(\ref{2nd_res}) in the form
\begin{equation}
  f(k,t)=\alpha(t)\e^{-2i\omega t}+
  \beta(t)\e^{-4i\omega t}+\mbox{h.c.}+\gamma(t)
  \label{2nd_ansatz}
\end{equation}
where we assume that $\alpha$, $\beta$ and $\gamma$ are slowly varying
functions that vary in a typical time scale of $A_0^{-4}$.  We also
expect that the coefficients of non--leading harmonics, $\beta$ and
$\gamma$, are suppressed by a factor of $A_0^2$ as compared to the
leading harmonics, $\alpha$.  Substituting the ansatz
(\ref{2nd_ansatz}) to eq.(\ref{2nd_res}) and set the coefficients of
$\e^{-2i\omega t}$, $\e^{-4i\omega t}$ and 1 to zero, one obtains the
following equations, respectively,
\begin{equation}
  -4i\dot{\alpha}+(\omega_k^2-4\omega^2+A_0^2)\alpha+
  {A_0^2\over2}\beta+{A_0^2\over2}\gamma+{A_0^4\over48}\alpha^*=0
  \label{linear_main}
\end{equation}
\[
  -12\beta+{A_0^2\over2}\alpha=0
\]
\begin{equation}
  4\gamma+{A_0^2\over2}(\alpha+\alpha^*)=0
  \label{5eq}
\end{equation}
From the last two equations one finds $\beta={A_0^2\over24}\alpha$,
$\gamma=-{A_0^2\over8}(\alpha+\alpha^*)$.  The first equation now can
be written as
\begin{equation}
  -4i\dot{\alpha}+(\omega_k^2-\omega_c^2)\alpha-{A_0^4\over24}\alpha^*
  =0
  \label{4i}
\end{equation}
where
$\omega_c^2=4\omega^2-A_0^2+{A_0^4\over24}=4+A_0^2+{A_0^4\over12}$.
This equation, except from the difference in numerical coefficients,
has the same form as that of the first resonance (eq.(\ref{1_res}).
As is this case, there are two solutions proportional to $\e^{s_kt}$
where
\begin{equation}
  s_k=\pm{1\over4}\sqrt{\left({A_0^4\over24}\right)^2-
  (\omega_k-\omega_c)^2}
  \label{sk}
\end{equation}
The resonance band is then
\[
  \omega_c^2-{A_0^4\over24}<\omega_k^2<\omega_c^2+{A_0^4\over24}
\]
Since $\omega_c$ is approximately $2$, the entire resonance band lies
entirely in the physical region. Some remarks are in order:

1.  The center of the resonance band is located at
$\omega_k=\omega_c\approx2+A_0^2/4$, i.e. is displaced by $A_0^2/4$
from 2.

2.  The width of the resonance band is of order $A_0^4$, and is much
smaller than the distance from the center of the resonance band to 2.
The rate of exponential enhancement $s_k$ is also of order $A_0^4$.

There may be resonance bands at higher $\omega_k$, however we will
ignore them since they are weaker than the one at $\omega_k\approx2$.

\subsection{Linear regime and its region of validity}

Now let us consider the behavior of the vacuum fluctuation on the
background $\phi_0$.  At $t=0$, the functions $f(k,t)$ are of order 1
for any value of $k$.  As $t$ increases, the modes inside the
resonance bands become large, while those lying outside the bands
remain small.  The most important are the modes in the strongest
physical band, i.e. the second one, which grows as $\e^{s_kt}$ where
$s_k$ is the positive value defined from eq.(\ref{sk}).  If $t\gg
A_0^{-4}$ only the modes close to the center of the band are
important.  From eq.(\ref{sk}) one finds for these modes
\begin{equation}
  f(k,t)\sim\exp\left({A_0^4\over96}t-
  {3(\omega_k-\omega_c)^2\over A_0^4}t\right)
  \label{Gaussian}
\end{equation}
and since the energy carried by the modes with momentum ${\bf k}$ is
proportional to $|f(k,t)|^2$, the energy distribution has the Gaussian
form centering at $\omega_k=\omega_c$ and with width of order
$t^{-1}$.  We will call the regime when eq.(\ref{Gaussian}) is valid
the linear regime, since the modes obey the linearized equation on the
background $\phi_0$.

Now let us find out the region of validity of this regime.  One may
expect that the linear regime is valid once the energy of the
inhomogeneous modes is much smaller than that of the homogeneous one.
However in fact the linear regime breaks down when the energy carried
by the inhomogeneous modes is still much smaller than the total
energy.  The reason for this is that the resonance band in our case
has a small width proportional to $A_0^4$, and a small change of the
amplitude of homogeneous oscillations can drive the resonance band to
a new location that does not overlap with the original resonance band.

Let us estimate the upper limit on the energy carried by
non--homogeneous modes for the linear approximation to be valid.  Due
to the energy conservation, when a portion of energy is transfered to
non--homogeneous modes, the amplitude of the homogeneous oscillation
decreases, and let us denote this as $A_0\to A_0-\delta A_0$.  The
center of the resonance band is located at $k=2+{A_0^2\over4}$, and we
required that this values is shifted by a value much smaller than the
width of the resonance band.  This leads to the condition
\[
  \delta A_0^2\ll A_0^4 \quad\mbox{or}\quad \delta A_0\ll A_0^3
\]
Since $\delta A_0/A_0$ is proportional to the fraction of energy
carried by inhomogeneous modes, the latter must be smaller than
$A_0^2$ of the total, for the linear regime to be valid,
\[
  \int\!{d{\bf k}\over(2\pi)^32\omega_k}\,|f(k,t)|^2\ll
  {A_0^4\over\lambda}
\]
Taking into account eq.(\ref{Gaussian}), one finds that the linear
regime breaks down at
\begin{equation}
  t\approx T_0={48\over A_0^4}\log{1\over\lambda A_0^2}
  \label{T_0}
\end{equation}
After $t=T_0$, the back--reaction of the created modes onto the
homogeneous oscillations cannot be neglected.  Note that at $t\sim
T_0$, the width of the Gaussian distribution, eq.(\ref{Gaussian}), is
$A_0^4\log^{-1/2}(1/\lambda A_0^2)$, i.e. much smaller than that of
the resonance band.

Before turning to Hartree--Fock approximation to go further than the
linear regime, let us make an important remark on the phase coherence
of the wave functions $f(k,t)$ of particles created during the linear
regime.  This coherence can be seen in the simple case of
$\omega=\omega_c$.  In this case $s_k={A_0^4\over96}$, and from
eq.(\ref{4i}) one sees that the exponentially increasing solution
corresponds to $\alpha$ having the phase $\pi/4$,
i.e. $\alpha=\alpha_0\e^{s_kt}\e^{i\pi/4}$ where $\alpha_0$ is a real
constant.  From eq.(\ref{2nd_ansatz}) one finds,
\[
  f(k,t)\sim\alpha_0\e^{s_kt}\cos\left(2\omega t-{\pi\over4}\right)
\]
Thus, except from the slowly varying exponential part, the wave
function $f(k,t)$ has a well--defined phase shift, namely $\pi/4$,
with $\phi_0^2(t)$.  It is worth noting that this phase shift is the
same for all vectors ${\bf k}$ pointing to all possible directions
that have $\omega_k=\omega_c$.  If one does not require that
$\omega_k=\omega_c$,, the phase shift is different, but since most
excited particles lies in a narrow central part of the resonance
band, this phase shift is approximately constant for all modes excited
during the linear regime.

\section{Hartree--Fock approximation}

\subsection{Hartree--Fock equations}

In the previous section we have seen that the linear regime is
violated rather early when the energy transfered to the inhomogeneous
modes is still small.  After the linear regime ceases to be valid, one
must take into account back-reaction of the inhomogeneous modes to the
homogeneous oscillations, in particular the decrease of the amplitude
of the latter due to the energy lost, and also the interaction between
the modes.  In this section, we will apply the Hartree--Fock
approximation to this problem (the Hartree--Fock approximation was
discussed earlier in connection with the problem of reheating in
Ref.\cite{Boyan,KLS}).  The region when this approximation is
justified will be discussed as well.

The Hartree--Fock equations has the form,
\[
  \left(\partial_t^2+1+{\lambda\over2}\langle\tilde{\phi}^2(t)\rangle
  \right)\phi_0(t)+{\lambda\over6}\phi_0^3(t)=0
\]
\begin{equation}
  \left(\partial_t^2+\omega_k^2+{\lambda\over2}\phi_0^2(t)+
  {\lambda\over2}\langle\tilde{\phi}^2(t)\rangle\right)f(k,t)=0
  \label{HF_eq}
\end{equation}
where $\phi_0$ and $\tilde{\phi}$ are the classical and quantum parts
of $\phi$, respectively.  Recalling eq.(\ref{quantiz}), the mean value
of $\tilde{\phi}^2$ is
\begin{equation}
  \langle\tilde{\phi}^2(t)\rangle=\int\!{d{\bf k}\over(2\pi)^32\omega_k}
  \,|f(k,t)|^2
  \label{26'}
\end{equation}
where $f$ is the mode function that is still to be found.  It is 
convenient to decompose $f$ into real and imaginary parts,
\[
  f(k,t)=F_1(k,t)+iF_2(k,t)
\]
where $F_1$ and $F_2$ are both real.  Eqs.(\ref{HF_eq}) can be
rewritten in the form
\begin{equation}
  \left(\partial_t^2+1+{\lambda\over2}\int\!
  {d{\bf q}\over(2\pi)^32\omega_q}\,\left(F_1^2(q,t)+F_2^2(q,t)\right)
  \right)\phi_0(t)+{\lambda\over6}\phi_0^3(t)=0
  \label{HF_eq1}
\end{equation}
\begin{equation}
  \left(\partial_t^2+\omega_k^2+{\lambda\over2}\phi_0^2(t)+
  {\lambda\over2}\int\!{d{\bf q}\over(2\pi)^32\omega_q}\,
  \left(F_1^2(q,t)+F_2^2(q,t)\right)\right)F_{1,2}(t)=0
  \label{HF_eq2}
\end{equation}
The initial conditions for these equations are formulated at $t=0$.
For $\phi_0$ these conditions are
\[
  \phi_0(0)={2\over\sqrt{\lambda}}\left(A_0+{A_0^3\over24}
  +O(A_0^5)\right), \qquad \dot{\phi}(0)=0
\]
while for $F_{1,2}$ one has from eq.(\ref{fekt})
\[
  F_1(k,0)=1,\qquad \dot{F}_1(k,0)=0,
\]
\begin{equation}
  F_2(k,0)=0,\qquad \dot{F}_2(k,0)=-2
  \label{F12_ini}
\end{equation}  
In analogy with the technique used in previous section, we will look
for the solution in the following form,
\[
  \phi_0(t)={1\over\sqrt{\lambda}}(A(t)\e^{-i\Omega}+
  B(t)\e^{-3i\Omega}+\mbox{h.c.})
\]
\begin{equation}
  F_i(k)={1\over\sqrt{\lambda}}(\alpha_i(k,t)\e^{-2i\Omega}+
  \beta_i(k,t)\e^{-4i\Omega}+\mbox{h.c.}+\gamma_i(k,t)), 
  \qquad i=1,2
  \label{HF_ansatz}
\end{equation}
where
\[
  \Omega(t)=\int\!dt\,\omega(t)
\]
and the functions $\omega(t)$, $A(t)$, $B(t)$, $\alpha_i(k,t)$,
$\beta_i(k,t)$, and $\gamma_i(k,t)$ are assumed to be slowly varying
functions of $t$.  Note that $\omega(t)$ can chosen so that $A(t)$ is
always real, in this case $\omega(t)$ has the meaning of the
time--dependent frequency of the $\phi_0$ oscillations.  Other
functions, i.e. $B$, $\alpha_i$, $\beta_i$ are, in general, complex.
The technique used here is similar to that applied to the case of
first and second resonance bands.  The calculation is rather
cumbersome, but straightforward (see Appendix A).  For our purpose, we
need only the equations determining the amplitudes of the main
harmonics of $\phi_0$ and $F_i(k,t)$, i.e. $A(t)$ and $\alpha_i(k,t)$.
The equation for $\alpha_i(k,t)$ is
\begin{equation}
  -4i\dot{\alpha}_i(k,t)+\left(\omega^2-\omega_c^2+{\cal I}(t)\right)
  \alpha_i(k,t)-\left({A_0^4\over24}-{\cal C}(t)\right)\alpha_i^*(k,t)=0
  \label{HF_main}
\end{equation}
where $i=1,2$, and
\[
  {\cal I}(t)=\int\!{d{\bf q}\over(2\pi)^32\omega_q}
  (|\alpha_1(q,t)|^2+|\alpha_2(q,t)|^2)
\]
\[
  {\cal C}(t)={1\over2}\int\!{d{\bf q}\over(2\pi)^32\omega_q}
  (\alpha_1^2(q,t)+\alpha_2^2(q,t))
\]
while $A(t)$ can be determined from $\alpha_i(k,t)$ by energy
conservation,
\[
  A^2(t)+4\int\!{d{\bf k}\over(2\pi)^32\omega_k}(|\alpha_1(k,t)|^2+
  |\alpha_2(k,t)|^2)=A_0^2
\]
Note that when ${\cal I}$ and ${\cal C}$ are still much smaller than
$A_0^4$, eq.(\ref{HF_main}) reduces to eq.(\ref{linear_main}), and the
linear regime is restored. It is useful to rescale
$\omega_k-\omega_c$, $t$ and $\alpha$ to expel the small parameter
$A_0$ from eq.(\ref{HF_main}).  Making use of the following change of
variables
\[
  \omega_k-\omega_c={A_0^4\over96}\kappa,\qquad
  \tau={A_0^4\over96}t
\]
\begin{equation}
  a_i(\kappa)={3^{1/4}\over4\pi}\alpha_i(k)
  \label{change_var}
\end{equation}
eq.(\ref{HF_main}) becomes the following simple equation for
$a(\tau,\kappa)$
\begin{equation}
  -i{\partial\over\partial\tau}a_i(\kappa)+(\kappa+I)a_i(\kappa)-
  (1-C)a_i^*(\kappa)=0
  \label{akappa}
\end{equation}
where now
\[
  I=\int\!d\kappa\,(|a_1(\kappa)|^2+|a_2(\kappa)|^2)
\]
\begin{equation}
  C={1\over2}\int\!d\kappa\,(a_1^2(\kappa)+a_2^2(\kappa))
  \label{29*}
\end{equation}
The initial conditions for $a_i(\kappa,t)$ can be found from
eqs.(\ref{F12_ini}), (\ref{HF_ansatz}) and (\ref{change_var}),
\[
  a_1(\kappa,0)={3^{1/4}\over8\pi}\sqrt{\lambda},\qquad
  a_2(\kappa,0)=-i{3^{1/4}\over8\pi}\sqrt{\lambda}
\]
It is also useful to write eq.(\ref{akappa}) in real notation.
Denoting the real and imaginary parts of $a_i$ as $u_i$ and $-v_i$,
respectively,
\[
  a_{1,2}=u_{1,2}-iv_{1,2}, \qquad i=1,2
\]
where $u_i$ and $v_i$ are real, the equations for $u_i$ and $v_i$ are
\[
  \dot{u}_i+(\kappa+I+1-\mbox{Re}C)v_i+\mbox{Im}Cu_i=0
\]
\begin{equation}
  \dot{v}_i-(\kappa+I-1+\mbox{Re}C)u_i-\mbox{Im}Cv_i=0
  \label{uv_eq}
\end{equation}
where
\[
  I=\int\!d\kappa\,(u_1^2+u_2^2+v_1^2+v_2^2)
\]
\begin{equation}
  \mbox{Re}C={1\over2}\int\!d\kappa\,(u_1^2+u_2^2-v_1^2-v_2^2),
  \qquad\mbox{Im}C=-\int\!d\kappa\,(u_1v_1+u_2v_2)
  \label{consistency}
\end{equation}
The initial conditions for $u_{1,2}$, $v_{1,2}$ are
\[
  u_1(\kappa,0)=v_2(\kappa,0)={3^{1/4}\over8\pi}\sqrt{\lambda},
 \qquad u_2(\kappa,0)=v_1(\kappa,0)=0
\]

\subsection{Limit $\log(1/\lambda)\to\infty$}

Let us briefly consider the limit when $\log(1/\lambda)$ is very
large.  More precisely, we consider the limit
$\log(1/\lambda)\to\infty$, while $t-T_0=\mbox{fixed}$ ($T_0$ is the
time moment when the linear regime breaks down, eq.(\ref{T_0})).  In
this limit, only modes close to the center of the resonance band are
important, so in eq.(\ref{akappa}) one can replace $\kappa$ with 0.
Then their solution has the form $u_i(\kappa)=U_if(\kappa)$,
$v_i=V_if(\kappa)$, where $f(\kappa)$ is arbitrary function of
$\kappa$, which will be normalized for convenience,
$\int\!d\kappa\,f^2(\kappa)=1$, and $U_i$ and $V_i$ do not depend on
$\kappa$ and satisfy the following equations,
\[
  \dot{U}_i+\left({1\over2}(U_1^2+U_2^2)+{3\over2}(V_1^2+V_2^2)+1
  \right)V_i-(U_1V_1+U_2V_2)U_i=0
\]
\begin{equation}
  \dot{V}_i-\left({3\over2}(U_1^2+U_2^2)+{1\over2}(V_1^2+V_2^2)-1
  \right)V_i+(U_1V_1+U_2V_2)U_i=0
  \label{UV_eq}
\end{equation}
Giving the initial conditions, this set of equation can be solved
numerically.  The typical time dependence of the total energy of
inhomogeneous modes, which is proportional to
$U_1^2+U_2^2+V_1^2+V_2^2$, is presented in Fig.1.  As one can see from
this figure, the energy of carrying by inhomogeneous modes first grows
exponentially, but at some time moment reach a maximum and then drops
back to 0.  Note that at the maximum the total energy of inhomogeneous
modes is of order $A_0^2$ of the total energy. If one follows
eqs.(\ref{UV_eq}) further, the energy will reach some very small
minimum and then the process would repeat again, however this occurs
outside the regime we are considering ($t-T_0=\mbox{fixed}$).

The reason why the energy of inhomogeneous modes goes back to the
homogeneous one after some time moment is that the behavior of the
system in this limit is the same as of the system of two coupled
oscillators.  In fact, imagine that the particle spectrum consists of
only modes with momenta ${\bf k}$ and $-{\bf k}$, then if one repeats
the whole calculation we have made one finds again eq.(\ref{UV_eq}).
The behavior shown in Fig.1 is typical for the system of two 
non--linearly coupled oscillators, where the energy is concentrated
at first in one oscillator and the frequency of the second one lies
in a resonance band on the background of the first.

\subsection{Finite $\log(1/\lambda)$.  Rate of particle production}

The regime discussed above ($\log(1/\lambda)\to\infty$,
$t-T_0=\mbox{fixed}$) is not, however, interesting from the physical
point of view.  In fact, even when the coupling $\lambda$ is small,
$\log(1/\lambda)$ is typically not very large.  On other hand, we are
interested in the behavior of the system up to thermalization, so
$t-T_0$ may be very large.  In these cases, the effect of the finite
width of the distribution of resonance modes is important, and the
system cannot be reduced to that of two coupled oscillators.

Since now the equation involves an infinite number of degrees of
freedom, it seems hard to find the exact solution analytically.  We
have made numerical simulations of the solution.  In Fig.2 the time
dependence of the total energy of inhomogeneous modes ($I$) is
presented at $\lambda=10^{-12}$.  As seen from this figure, at the
beginning the energy of inhomogeneous modes is very small by it
increases exponentially.  At some time moment, however, it reaches
some maximum, and the energy decreases.  It does not reach 0 however,
since $\log(1/\lambda)$ is finite.  The energy makes some more
oscillations, and subsequent evolution is rather complicated, but
shows a clear tendency to grow.

Fortunately, it is possible to find analytically the asymptotics of
$I$ in the limit of large $\tau$, if one makes some reasonable
assumptions on the behavior of $I(\tau)$ and $C(\tau)$.  Let us
suggest that $I(\tau)$ and $C(\tau)$ are smooth, slowly varying
functions of $\tau$ at large $\tau$.  We assume that the imaginary
part of $C$ is very small and can be neglected, and $\mbox{Re}C<1$.
The last assumption is that $I(\tau)+C(\tau)$ and $I(\tau)-C(\tau)$
are monotone, increasing functions of $\tau$.  To find the
asymptotics of $I(\tau)$ and $C(\tau)$, we will try to solve
eq.(\ref{akappa}) at each $\kappa$ and demand latter that the
conditions (\ref{29*}) are satisfied, so that our calculations are
self--consistent.  For the reason that will be clear in further
discussions, we are interested only in large negative values of
$\kappa$.

First assume that $I$ and $C$ are constant.  Then eq.(\ref{akappa})
has the same form as eq.(\ref{linear_main}), except that the center of
the resonance band is now located at $\kappa=I$ and its width is
$1-C$.  Now let $I$ and $C$ change smoothly with $\tau$.  Then for a
given value of $\kappa$ one can determine the two time moments
$\tau_1$ and $\tau_2$ which are the solutions to the equations
$I+(1-C)=\kappa$ and $I-(1-C)=\kappa$, respectively.  The evolution of
the mode $a(\kappa)$, thus can be divided into 3 epochs,

1.  $0<\tau<\tau_1$, where $-\kappa>I-(1-C)$.  During this time
interval, $\kappa$ lies outside the resonance band.  In this case
$u_i(\tau)$ and $v_i(\tau)$ are oscillatory and the amplitudes of
$a_i(\kappa)$ remain small.

2.  $\tau_1<\tau<\tau_2$, where $\kappa$ lies inside the resonance
band.  This is the time interval when $a_i(\kappa)$ is enhanced by a
large factor due to parametric resonance.  To estimate this factor let
us denote
\[
  s_\kappa(\tau)=\sqrt{(1-C)^2-(\kappa+I)^2}
\]
to be the time--dependent rate of exponential growth, and if
$s_k(\tau)$ is slowly changing the enhancement factor during the whole
period from $\tau=\tau_1$ to $\tau=\tau_2$ is proportional to
\[
  \exp\left(\int\!d\tau\,s_\kappa(\tau)\right)=\exp\left(
  {\pi\over2}\cdot{(1-C)^2\over\dot{I}}\right)
\]
where $\dot{I}=\partial I/\partial\tau$ we have assumed that $\dot{I}$
and $C$ can be considered as constant when $\tau$ varies from $\tau_1$
to $\tau_2$.  Since this enhancement factor must be of order
$\lambda^{-1/2}$ for $a_i$ to increase from $\sqrt{\lambda}$ to 1, one
obtains the following equation,
\begin{equation}
  \dot{I}={\pi(1-C)^2\over\log{1\over\lambda}}
  \label{Idot}
\end{equation}
which is valid with the logarithmic precision.

3.  $\tau>\tau_2$, when $\kappa$ is again outside the resonance band.
Now $u$ and $v$ are again oscillatory and satisfy the equation
\[
  \dot{u_i}+(\kappa+I+(1-C))v_i=0
\]
\begin{equation}
  \dot{v_i}-(\kappa+I-(1-C))u_i=0
  \label{3rd}
\end{equation}
Making use of the fact that $I$ and $C$ are slowly varying functions,
the solution to eqs.(\ref{3rd}) has the form
\[
  u_i=\lambda_i(\tau)\sqrt{\kappa+I+(1-C)}\cos\left(\varphi_i(\tau)\right)
\]
\begin{equation}
  v_i=\lambda_i(\tau)\sqrt{\kappa+I-(1-C)}\sin\left(\varphi_i(\tau)\right)
  \label{uvlambda}
\end{equation}
where
\[
  \varphi_i(\tau)=\int\limits_{\tau_2}^\tau\!d\tau'\,
  \sqrt{(\kappa+I(\tau'))^2-(1-C(\tau'))^2}+\varphi_i(0)
\]
and $\lambda_i(\tau)$ are slowly varying functions, and $\varphi_i(0)$
are some constants that will not be important in our further
calculations.  The time dependence of $\lambda(\tau)$ can be found,
making use of the adiabatic invariance technique \cite{Landau}.  The
results reads
\[
  \lambda_i(\tau)={\lambda_i(\kappa)\over((\kappa+I)^2-(1-C)^2)^{1/4}}
\]
where $\lambda_i(\kappa)$ does not depend on $\tau$.  So, we find the
solution to eq(\ref{3rd}) for given functions $I(\tau)$ and $C(\tau)$,
\[
  u_i=\lambda_i(\kappa)\left({\kappa+I+(1-C)\over\kappa+I-(1-C)}\right)
  ^{1/4}\cos\left(\varphi_i(\tau)\right)
\]
\[
  v_i=\lambda_i(\kappa)\left({\kappa+I-(1-C)\over\kappa+I+(1-C)}\right)
  ^{1/4}\sin\left(\varphi_i(\tau)\right)
\]
Now we have to satisfy the self--consistency conditions,
eqs.(\ref{consistency}).  At large $\tau$ one has $I\gg(1-C)$ and both
integrals are dominated by the region $-I+(1-C)<\kappa<0$.  The first
self--consistency condition reads,
\[
  I=\int\limits_{-I+(1-C)}^0\!d\kappa\,
  \left(\sqrt{\kappa+I+(1-C)\over\kappa+I-(1-C)}
  \left(\lambda_1^2(\kappa)\cos^2(\varphi_1(\tau))+
  \lambda_2^2(\kappa)\cos^2(\varphi_2(\tau))\right)+\right.
\]
\[
  \left.\sqrt{\kappa+I-(1-C)\over\kappa+I+(1-C)}
  \left(\lambda_1^2(\kappa)\sin^2(\varphi_1(\tau))+
  \lambda_2^2(\kappa)\cos^2(\varphi_2(\tau))\right)\right)
\]
Now let us note that as $\cos(\varphi_i)$ and $\sin(\varphi_i)$ are
rapidly oscillating functions, one can replace $\cos^2(\varphi_i)$ and
$\sin^2(\varphi_i)$ by $1/2$ and obtains
\begin{equation}
  I\approx\int\limits_{-I+(1-C)}^0\!d\kappa\,
  (\lambda_1^2(\kappa)+\lambda_2^2(\kappa))
  \label{Iint}
\end{equation}
Since this equation must be valid for any large value of $\tau$, or
equivalently, for any large negative value of $-I+(1-C)$, one finds
$\lambda_1^2(\kappa)+\lambda_2^2(\kappa)=1$ for every $\kappa$,
$|\kappa|\gg1$.

The second condition then reads,
\begin{equation}
  C={1\over4}\int\!d\kappa\,\left(
  \sqrt{\kappa+I+(1-C)\over\kappa+I-(1-C)}-
  \sqrt{\kappa+I-(1-C)\over\kappa+I+(1-C)}\right)
  \label{Cint}
\end{equation}
The integral in the r.h.s.\ can be evaluated to the logarithmic
accuracy, and as the result one obtains the following equation,
\begin{equation}
  C={1-C\over2}\log{I\over1-C}
  \label{CI}
\end{equation}
Note that both the integrals in (\ref{Cint}) and (\ref{Iint}) are
dominated by large negative values of $\kappa$ that we are interested
in from the beginning.  From eqs.(\ref{Idot}) and (\ref{CI}), one
finds the following asymptotics of $I(\tau)$ and $C(\tau)$ in the
limit $\tau\to\infty$,
\begin{equation}
  I(\tau)\sim{4\pi\over\log{1\over\lambda}}\cdot{\tau\over\log^2\tau},
  \qquad 1-C\sim{2\over\log\tau}
  \label{ICt}
\end{equation}
So, at large $\tau$, $I(\tau)$ grows almost linearly, except from a
logarithmic correction.  In further calculations we will be interested
mostly in the order of magnitude of physical quantities, and in all
estimations we will neglect the logarithms (i.e. we consider
$\log\lambda^{-1}$ and $\log A_0^{-1}$ as quantities of order 1).
Recall that $I(\tau)$ is proportional to the density of particles
created by parametric resonance, and restore all power of $A_0$ from
eq.(\ref{change_var}), one can sees that eq.(\ref{ICt}) means that at
$t\gg A_0^{-4}$ the rate of particle creation is of order
$\lambda^{-1}A_0^8$.  Note that this rate is larger than that predicted
by the elementary theory, eq.(\ref{dn0/dt}), by a factor of
$1/\lambda$ (so one has $\dot{n}_0\sim\lambda^2n_0^4$ instead of
$\lambda^4n_0^4$.

To summarize, we find that in the long run the resonance band moves
toward lower energies ($I$ increases with $t$ and the center of the
resonance band is located at $\kappa=-I$).  Due to this movement,
particles with lower and lower energies are created, which in its turn
causes the resonance band to move to smaller energies.  The width of
the resonance band decreases during this process (as $\log^{-1}\tau$).
It is worth noting, however, that the width of the particle spectrum,
though much larger than that of the initial resonance band,
i.e. $A_0^4$, is still much smaller than 1, so all created particles
have energy close to 2.  The excitation of a given mode by parametric
resonance is a coherent process, but one the parametric enhancement
terminates, the phase of the mode become rapidly uncorrelated with
that of the homogeneous oscillations and of the other modes.  At $t\gg
A_0^{-4}$, most of the modes that have been excited lie outside the
resonance band and can be considered as ordinary particles.

It is useful to find the distribution function of the particles
produced by this mechanism.  This can be done by comparing 
eq.(\ref{26'}) with the one giving the fluctuation 
$\langle\tilde{\phi}^2\rangle$ for the state characterized by the
distribution $n_{\bf k}$,
\[
  \langle\tilde{\phi}^2\rangle=\int\!{d{\bf k}\over(2\pi)^32\omega_k}
  (2n_{\bf k}+1)
\]
If $n_{\bf k}\gg1$ (as will be obtained below), one finds
\[
  n_{\bf k}={1\over2}\langle|f(k,t)|^2\rangle
\]
where the average in the r.h.s.\ can be taken over time or over a finite
volume in the ${\bf k}$--space.   Now consider some mode that the 
resonance band has gone by.  At this mode one has
$u_1^2+v_1^2+u_2^2+v_2^2=\lambda_1^2+\lambda_2^2=1$, which, according
to eq.(\ref{change_var}), implies
$|\alpha_1(k)|^2+|\alpha_2(k)|^2=16\pi^2/\sqrt{3}$.  The occupation
number is then
\[
  n_{\bf k}(t)={1\over2}\langle|f(k,t)|^2\rangle=
  {1\over\lambda}\left(|\alpha_1|^2+|\alpha_2|^2\right)=
  {16\pi^2\over\sqrt{3}}\cdot{1\over\lambda}
\]
Therefore, at given time moment $t\gg A_0^{-4}$, the particle
distribution has the form of a thin spherical shell.  The outer sphere
of this shell is fixed at $\omega_k=\omega_c$, while the inner sphere
moves inward when $t$ increases.  The occupation number of modes
inside the shell is of order $1/\lambda$.

Now let us estimate when the number of particle with energy $E=2$ when
their scattering cannot be neglected.  Consider the scattering process
where two particles with energy $E=2$ scatter to one particle in the
condensate (i.e. with energy 1) and another particle with energy
$E=3$.  This process is in fact the most probable since it involves
one particle in the condensate, and thus has a large
Bose--enhancement factor associated with the latter.  The criterion
for the scattering to be important is that the rate of production of
particles with energy $E=3$ is comparable with those with energy $E=2$.
Let us suppose that the density of the particle with energy $E=2$ is
$n$.  Then the number of particle with energy $E=3$ created by the
scattering process in an unit volume during an unit time interval is
$\sigma n^2n_0$, where $\sigma$ is the cross section and $n_0$ is the
density of the condensate.  Substituting $\sigma\sim\lambda^2$ and
$n_0\sim A_0^2/\lambda$, one finds that this rate is $\lambda
A_0^2n^2$.  Compare with the rate of creation of particles with energy
$E=2$, which is of order $\lambda^{-1}A_0^8$, one finds that the
scattering becomes important when $n\sim\lambda^{-1}A_0^3$, which
occurs at the time moment $t\sim A_0^{-5}$.  Note that at $t\sim
A_0^{-5}$ the total energy of inhomogeneous modes is
$\lambda^{-1}A_0^3$, which is still much smaller than the total energy
($\lambda^{-1}A_0^2$).

\section{Particle scattering and formation of Kolmogorov turbulence 
spectrum}

In the previous section we have found that the Hartree--Fock
approximation breaks down at $t\sim A_0^{-5}$, when most of the energy
is still carried by the homogeneous mode.  In this section we will
consider the behavior of the system after the moment $t\sim A_0^{-5}$
up to the time when the the homogeneous mode begins to lost a
considerable portion of its energy.  Since the energy flux from the
homogeneous mode is $\lambda^{-1}A_0^8$ and the total energy is of
order $\lambda^{-1}A_0^2$, the latter time moment is $A_0^{-6}$.
Thus, in this section we consider $A_0^{-5}\ll t\ll A_0^{-6}$.

The system can be characterized by the distribution function $n_{\bf
k}(t)$, while the homogeneous oscillations correspond to the
delta--functional part of $n_{\bf k}$ at ${\bf k}=0$.

Let us first write down the generic Boltzmann equation (the Boltzmann
equation for the $\phi^4$ field theory has been studied in a different
setting, see Refs.\cite{Semikoz1,Semikoz2}).  Since most of our
discussion will be based on order--of--magnitude estimations, we will
neglect all numerical factors.  One writes,
\[
  {dn_{\bf k}\over dt}\propto\lambda^2
  \int\!{d{\bf k}_1d{\bf k}_2d{\bf k}_3\over
  \omega_k\omega_{k_1}\omega_{k_2}\omega_{k_3}}\,
  (n_{{\bf k}_1}n_{{\bf k}_2}(1+n_{{\bf k}})(1+n_{{\bf k}_3})-
  n_{\bf k}n_{{\bf k}_3}(1+n_{{\bf k}_1})(1+n_{{\bf k}_2}))
\]
\[
  \delta({\bf k}_1+{\bf k}_2-{\bf k}_3-{\bf k})
  \delta(\omega_{k_1}+\omega_{k_2}-\omega_{k_3}-\omega_k)
\]
In fact, one should take into account the particle creation by
parametric resonance, however let us first neglects this effect.
Since the occupation numbers are typically of order $1/\lambda$ which
are much larger than 1, one can neglect the terms 1 compared to $n$.
The Boltzmann equation now becomes
\[
  {dn_{\bf k}\over dt}\propto
  \lambda^2\int\!{d{\bf k}_1d{\bf k}_2d{\bf k}_3\over
  \omega_k\omega_{k_1}\omega_{k_2}\omega_{k_3}}\,
  \left(n_{{\bf k}_1}n_{{\bf k}_2}n_{{\bf k}_3}+
   n_{{\bf k}_1}n_{{\bf k}_2}n_{\bf k}-
   n_{\bf k}n_{{\bf k}_3}n_{{\bf k}_1}-
   n_{\bf k}n_{{\bf k}_3}n_{{\bf k}_2}\right)
\]
\begin{equation}
  \delta({\bf k}_1+{\bf k}_2-{\bf k}-{\bf k}_3)
  \delta(\omega_1+\omega_2-\omega_k-\omega_3)
  \label{nnn}
\end{equation}
Let us make two remarks on eq.(\ref{nnn}).  First, since we have
$n_{\bf k}\sim\lambda^{-1}$, the time scale set by the Boltzmann
equation does not depend parametrically on $\lambda$ (but may depend
on $A_0$).  Second, if at some $t$ the mode ${\bf k}$ is unoccupied,
$n_{\bf k}=0$, the particles will be created at this ${\bf k}$ only
when there is a process ${\bf k}_1,{\bf k}_2\to{\bf k}_3,{\bf k}$
where the other three momenta participating in the process have
nonzero occupation numbers.  Since the initial condition for the
Boltzmann equation is that where all particles have energy $E=1$
or 2, approximately, one can expect that the particle distribution
will have peaks at $E=3$, 4... i.e. at integer values of $E$
when $t\sim A_0^{-5}$.  In fact, if three particles participating in a
scattering have integer energies, the energy of the fourth is also
integer due to energy conservation.  This ``peak--like'' structure of
the spectrum will not preserve infinitely long: since the particles
produced by parametric resonance have not exactly the same energy, the
width of the peaks will become broader and broader, and at some time
moment comparable to 1, however this happens at some time scale much
larger than $A_0^{-5}$.

Therefore, we take the distribution function in the form
\begin{equation}
  n_{\bf k}=n_0\delta({\bf k})+
  \sum_{l=2}^\infty{n_l\over4\pi k_l^2}\delta(\omega_k-l)
  \label{Boltz_ansatz}
\end{equation}
where we have introduced the notation $k_l=\sqrt{l^2-1}$, which is the
momentum corresponding to the energy $\omega_k=l$.  Note that $n_l$ has
the meaning of the density of particles with energy $l$.
Before substituting eq.(\ref{Boltz_ansatz}) to eq.(\ref{nnn}), let us
also notice that the scattering processes ${\bf k}_1,{\bf k}_2\to{\bf
k}_3,{\bf k}$ (where ${\bf k}\neq0$) can be divided into 2 types:
those without participation of condensate particles (i.e. all ${\bf
k}_1,{\bf k}_2,{\bf k}_3$ are nonzero), and those with one condensate
particle participating in the scattering (one of ${\bf k}_1,{\bf
k}_2,{\bf k}_3$ is 0).  In order to avoid the overlap with the
Hartree--Fock effects, it is necessary to exclude from consideration
the scattering where two of the participating particles have zero
momenta.  It is natural to expect that when the density of
non--condensate particles is still small, the processes of the second
type (which will be called ``three--particle'' processes) dominate over
those of the first type (``four--particle'' transitions).  The criterion
for the 4--particle scattering to be negligible compared to the
3--particles ones can be easily found.  In fact, the 3--particle
processes are enhanced by the Bose factor of $n_0$ due to the
participation of a condensate particle in the scattering, while the
4--particle processes contain instead of this factor an additional
momentum integral $\int\!d{\bf k}\,n_{\bf k}/\omega_k$.  So, the
criterion for the 4--particle collision to be negligible is
\begin{equation}
  n_0\gg\int\!{d{\bf k}\over\omega_k}n_{\bf k}
  \label{4part}
\end{equation}
Running ahead, let us state that this condition is satisfied in the
whole evolution up to the moment when thermalization is completed.

At first, one can consider the density of particles in the condensate
$n_0$ as a constant.  Substituting eq.(\ref{Boltz_ansatz}) to the
Boltzmann equation, and throwing away terms corresponding to
four--particle collisions, after some calculations (see Appendix B)
the following equation is obtained,
\begin{equation}
  {d n_l\over dt}\propto\lambda A_0^2\left({1\over2}\sum_{l_1+l_2=l+1}
  {n_{l_1}n_{l_2}\over k_{l_1}k_{l_2}l_1l_2}+
  \sum_{l_1-l_2=l-1}{n_{l_1}n_{l_2}\over k_{l_1}k_{l_2}l_1l_2}
  -{n_l^2\over k_l^2l^2}-{2n_l\over k_ll}\sum_{l_1<l}{n_l\over k_ll}
  \right)
  +j\delta_{l2}
  \label{nn}
\end{equation}
where we have included
into the r.h.s.\ the term $j\delta_{l2}$ to take into account the
production of particles with energy closed to $E=2$ that we have
found in the Hartree--Fock approximation.  Therefore, the Boltzmann
equation is now written in terms of a discrete set of numbers $n_l$
which is proportional to the density of particles with energy closed to
$l$.  Let us emphasize that in eq.(\ref{nn}) we do not write down the
numerical constants.  For estimation, it is sufficient to take a
constant $j\sim\lambda^{-1}A_0^8$, although in fact we found in Sect.3
that the production rate is slowly changing.

As seen from eq.(\ref{nn}), due to the scattering, particles with
energy $E=3$ are created by colliding two particles with energy $E=2$.
Then particles with energy $E=4$ and 5 come to existence in the
collision of particles with energy $E=2$ and 3, etc.  Therefore,
particles with higher and higher energies are created, despite the
fact that the source term in eq.(\ref{nn}) produces particles with
$E=2$ only.

The situation when energy is transfered from the long--distance modes
to short--distance ones is the common situation in turbulent systems
(see, for instance, \cite{Zakharov}).  Hence one can make use of the
vast intuition elaborated in the theory of turbulence in order to
predict the long--term behavior of the system described by
eq.(\ref{nn}).  In particular, in the turbulent systems, if one has a
constant flow of energy into the modes at some momentum scale during
an infinite time interval, a stationary distribution is formed.  To
find this distribution in our case, one looks for those which has
vanishing collision integral,
\begin{equation}
  {1\over2}\sum_{l_1+l_2=l+1}
  {n_{l_1}n_{l_2}\over k_{l_1}k_{l_2}l_1l_2}+
  \sum_{l_1-l_2=l-1}{n_{l_1}n_{l_2}\over k_{l_1}k_{l_2}l_1l_2}
  -{n_l^2\over k_l^2l^2}-{2n_l\over k_ll}\sum_{l_1<l}{n_l\over k_ll}
  +\lambda^{-1}A_0^6\delta_{l2}=0
  \label{stationary}
\end{equation}
While eq.(\ref{stationary}) seems hard to solve analytically, the
asymptotics of $n_l$ at large $l$ can be found.  In analogy with the
Kolmogorov spectrum of turbulence, the stationary solution can be
expected to have the form $n_l\sim l^\eta$ at large $l$.  For
convenience, it is useful to introduce
\[
  n(E)={n_l\over4\pi l^2},\qquad l\approx E
\]
which (at large $E$) has the meaning of the average occupation number
$n_{\bf k}$ at energy $\omega_k=E$ (one can imagine $n(E)$ as the
distribution function after one smears off the peaks of the spectrum).  
At large $l$, the sums in eq.(\ref{nn}) can be transformed to 
integrals, and one obtains,
\[
  {dn(E)\over dt}\propto\lambda A_0^2E^{-2}\left(
  {1\over2}\int\limits_0^E\!dE_1\,n(E_1)n(E-E_1)+
  \int\limits_0^\infty\!dE_1\,
  n(E_1)n(E+E_1)-\right.
\]
\[
  \left.-2n(E)\int\limits_0^E\!dE_1\,n(E_1)\right)=0
\]
Substituting the ansatz $n(E)\sim E^\eta$ to this equation, one
obtains the following equation on the Kolmogorov index $\eta$ for
$n(E)$ to be stationary,
\begin{equation}
  {1\over2}\int\limits_0^1\!dx\,x^\eta(1-x)^\eta+
  \int\limits_0^\infty\!dx\,x^\eta(1+x)^\eta-
  2\int\limits_0^1\!dx\,x^\eta=0
  \label{alpha_eq}
\end{equation}
The l.h.s.\ is finite when $\eta$ lies in the region $-3<\eta<-1/2$.
It can be verified that eq.(\ref{alpha_eq}) has exactly two solutions,
$\eta=-1$ and $\eta=-3/2$ (accidentally, these values coincide with
the Kolmogorov indices in the case of non--relativistic Bose gas with
condensate, see Ref.\cite{Semikoz2}).  The physical meaning of the
first solution is trivial: it corresponds to the thermal distribution
$n_{\bf k}\sim\omega_k^{-1}$, and is not appropriate for our purposes.
The second solution is a non--trivial one, $n_{\bf
k}\sim\omega_k^{-3/2}$, and describes the particle distribution
developed in our system after an infinite time interval.

In fact, this stationary spectrum can be established only if one have
the energy flux to the mode with $E=2$ during an infinitely long time
period (this can be seen from the fact that the stationary spectrum
$n_{\bf k}\sim\omega_k^{-3/2}$ has infinite energy).  In the real
situation when this time interval is finite, one expects that the
Kolmogorov spectrum is cut off in the ultraviolet.  In other words, at
finite $t$ one has $n_{\bf k}\sim\omega_k^{-3/2}$ at $k\ll k_0(t)$ and
$n_{\bf k}\approx0$ at $k\gg k_0(t)$.  The cutoff $k_0$ is expected to
increase with $t$.  Note that for this type of spectrum both energy
and particle number is concentrated at the cutoff, $k\sim k_0$.

That the Kolmogorov index is equal $-3/2$ can be understood in a very
intuitive way.  Suppose that at some time moment the particle
distribution has the form of $n_{\bf k}\sim\omega_k^\eta$, with a
cutoff at $k_0$.  If $\eta$ is the correct Kolmogorov index, the
cutoff $k_0$ increases with time, while the occupation numbers at
$k\ll k_0$ do not change.  The time scale during which the cutoff is,
say, doubled, is of the same order as the time scale set by the
Boltzmann equation, which can be estimated to be proportional to
$k_0/n(k_0)\sim k_0^{1-\eta}$.  During this time interval, an amount
of energy, also proportional to $k_0^{1-\eta}$ is fed into the
system, which is spent to the creation of particles with energies
between $k_0$ and $2k_0$.  The total energy of these particles is
$n(k_0)k_0^4\sim k_0^{4+\eta}$.  So, we obtain the equation
$1-\eta=4+\eta$, which gives the Kolmogorov index $\eta=-3/2$.

The power law $n_{\bf k}\sim\omega_k^{-3/2}$ is valid only at $k\ll
k_0$. Let us consider the particle spectrum at the momentum scale of
the cutoff, $k\sim k_0$.  In analogy with the situation in turbulence
systems \cite{Zakharov,Semikoz1,Semikoz2}, one can expect that the
distribution has the following self--similar behavior,
\begin{equation}
  n_{\bf k}(t)={c_1\over t^q}f\left(c_2k\over t^p\right)
  \label{self_sim}
\end{equation}
where $c_1$ and $c_2$ are some constants chosen so that the function
$f(\xi)$ varies on the typical values of $\xi$ of order 1 and is
itself of order 1.  We will try to find $p$ and $q$ but will not
attempt to obtain the form of the function $f(\xi)$.  Substituting
this ansatz to the Boltzmann equation, and comparing the time
dependence of the left and right hand sides, one finds that the latter
can be satisfied only if the following relation between $p$ and $q$
holds,
\begin{equation}
  p+q=1
  \label{p+q=1}
\end{equation}
On the other hand, one expects that when $k\ll k_0$ eq.(\ref{self_sim})
must reproduce the Kolmogorov distribution.  This fact implies
that the asymptotics of the function $f(\xi)$ at small $\xi$ is
$f(\xi)\sim\xi^{-3/2}$.  Furthermore, in the same region $k\ll k_0$,
the spectrum must be stationary, i.e. $n_{\bf k}$ does not depend on
$t$.  Taking into account these two facts, one obtains one more
equation on $p$ and $q$,
\begin{equation}
  {3p\over2}=q
  \label{3p/2=q}
\end{equation}
Solving eqs.(\ref{p+q=1}) and (\ref{3p/2=q}), we find $p=2/5$,
$q=3/5$.  In particular the cutoff $k_0$ grows with time as $k_0\sim
t^{2/5}$.  The total energy of the system behaves like 
$c_1c_2^{-4}t^{4p-q}\sim c_1c_2^{-4}t$, which agrees with the fact 
that the energy is fed into the
system with a constant rate.  Since the rate of creation of particles
with energy close to 2 is $\lambda^{-1}A_0^8$, one obtains
$c_1c_2^{-4}=\lambda^{-1}A_0^8$.  On the other hand, at $t\sim A_0^{-5}$,
the typical momentum of particles is 1, so $c_2=A_0^{-2}$.  Therefore,
$c_1=\lambda^{-1}$, and eq.(\ref{self_sim}) looks like the following,
\begin{equation}
  n_{\bf k}(t)={1\over\lambda}\cdot{1\over t^{3/5}}f\left(
  {k\over A_0^2t^{2/5}}\right)
  \label{self_sim1}
\end{equation}

Let us emphasize that in our calculations are limited to the range
$A_0^{-5}\ll t\ll A_0^{-6}$.  It is straightforward to verify the
condition for the 4--particle collisions to be negligible: the l.h.s.\
of eq.(\ref{4part}) is of order $\lambda^{-1}A_0^2$ while the r.h.s.\ 
is always smaller than $A_0^{14/5}$ in the time interval we are 
considering, $A_0^{-5}\ll t\ll A_0^{-6}$.

Another note is in order.  The total kinetic energy of the produced
particles grow almost linearly with time, but since the average energy
of the particles also grows as $t^{2/5}$, the number of particles
grows only as $t^{3/5}$.  This means that the larger is $t$, the
slower is the growth of the total number of particles outside the
condensate.  This slowdown occurs despite the fact that particles with
energy $E=2$ are produced with approximately constant rate.  There is
in fact no contradiction since the dominant collision process in our
case is the $2\to2$ scattering where two particles with nonzero
momenta collide and become one particles in the condensate and one
particle with nonzero momentum: each such act effectively reduces the
number of particles with nonzero momentum by 1.  At large $t$ most
particles produced at the energy level 2 effectively go back to the
condensate by this mechanism, transferring their kinetic energy to the
particles that are already outside the condensate.  The condensate
looses 4 particles in each $4\to2$ transition but gains back 2
particles from the scattering, therefore the actual rate of condensate
evaporation is smaller than predicted by the Hartree--Fock
approximation by a factor of 2.

For further reference let us note that at the end of the epoch that we
are considering, $t\sim A_0^{-6}$, the typical momentum of particles
is $A_0^{-2/5}$ and the typical occupation number is
$\lambda^{-1}A_0^{18/5}$.  Since now the typical energy is much larger
than 1, one can neglect the peak structure of the spectrum and deal
with the average occupation number $n(E)$ only, regardless how
effective the processes broadening the peaks are.

\section{Thermalization}

\subsection{Early step of thermalization: $A_0^{-6}\ll t\ll A_0^{-8}$}

When $t$ become much larger than $A_0^{-6}$, the energy carried by the
condensate becomes much smaller than the total.  Therefore, one can
consider the regime $t\gg A_0^{-6}$ as the epoch of thermalization of
particles produced by condensate evaporation.  Though the condensate
now carries only a small fraction of the energy, it still plays an
important role in the scattering process.  In fact, we will find that
the 3--particle collisions are still dominant.  Let us first find out
the time dependence of the condensate density.  To do that let us
first assume that the the dominant mechanism of the decay of the
condensate density is the coherent particle creation via parametric
resonance.  Then the decay of $n_0$ is describes by the equation
$\dot{n}_0=\lambda^3n_0^4$, from which one finds
\[
  n_0\sim \lambda^{-1}t^{-1/3}, \qquad t\gg A_0^{-6}.
\]
The Boltzmann equation, where only 3--particle collisions are taken
into account, now has the form
\[
  {dn(E)\over dt}\propto\lambda t^{-1/3}E^{-2}\left(
  {1\over2}\int\limits_0^E\!dE_1\,n(E_1)n(E-E_1)+
  \int\limits_0^\infty\!dE_1\,n(E_1)n(E+E_1)\right.
\]
\begin{equation}
  \left.-2n(E)\int\limits_0^E\!dE_1\,n(E_1)\right)
  \label{Boltz_1/3}
\end{equation}
It is natural to assume that the particle distribution still shows a
self--similar behavior (\ref{self_sim}) in this regime with a some
values of the parameters $p$, $q$, $c_1$ and $c_2$.  Substituting the
ansatz (\ref{self_sim}) into eq.(\ref{Boltz_1/3}) one obtains the
following relation between $p$ and $q$,
\[
  p+q={2\over3}
\]
On the other hand, the energy should be conserved (since most energy
has been emitted from the condensate at the time scale $t\sim
A_0^{-6}$), which leads to the equation $q=4p$.  Therefore, $p=2/15$ and
$q=8/15$.  To find $c_1$ and $c_2$ one makes use of the matching
condition with the previous regime at $t\sim A_0^{-6}$ where we have
found $k_0\sim A_0^{-2/5}$ and $n_{\bf k}\sim\lambda^{-1}A_0^{18/5}$.  
We obtains
$c_1=\lambda^{-1}A_0^{2/5}$ and $c_2=A_0^{-2/5}$, and the self--similar
behavior of $n_{\bf k}$ on $t$ has the form
\begin{equation}
  n_{\bf k}(t)={1\over\lambda}\cdot{A_0^{2/5}\over t^{8/15}}
  f\left({k\over A_0^{2/5}t^{2/15}}\right)
  \label{2/15}
\end{equation}

There are two assumptions that have been used in order to find
eq.(\ref{2/15}) that we now have to check.  The first is that
particles in the condensates are lost mainly due to the parametric
resonance condensate evaporation.  To find when this assumption is
justified, let us mention that the concurrent process that changes the
density of the condensate is the particle collision: as noted above,
the dominant scattering is supposed to be those where 2
non--condensate particles collide and becomes one non--condensate
particle and a condensate one and the number of particle in the
condensate is increased by 1.  Therefore, there is a flux of particles
coming from the non--condensate modes back to the condensate, which
can be calculated by taking the time derivative of the total number of
non--condensate particles.  We will refer to this phenomenon as ``Bose
condensation'' (cf. Refs.\cite{Semikoz1,Semikoz2}). 

According to eq.(\ref{2/15}), the total number of particles with
nonzero momentum is $\lambda^{-1}A_0^{8/5}/t^{2/15}$, so the flux of
particles to the condensate is $\lambda^{-1}A_0^{8/5}/t^{17/15}$.  On
the other hand, the rate of particle creation by parametric resonance
is $n_0^4\sim \lambda^{-1}t^{-4/3}$.  Comparing the two quantities,
one sees that the former process can be neglected when $t\ll
A_0^{-8}$.  The Bose condensation becomes comparable to condensate
evaporation at $t\sim A_0^{-8}$.  So, the region of validity of
eq.(\ref{2/15}) is $A_0^{-6}\ll t\ll A_0^{-8}$.

Another requirement that must be verified is eq.(\ref{4part}).  The
l.h.s.\ is $\lambda^{-1}t^{-1/3}$ while the r.h.s.\ is of order
$\lambda^{-1}A_0^{6/5}/t^{4/15}$.  It is easy to see that the l.h.s.\ 
is larger than the r.h.s.\ everywhere in the region 
$A_0^{-6}\ll t\ll A_0^{-8}$.

According to eq.(\ref{2/15}), at the end of the regime, $t\sim
A_0^{-8}$, the typical energy of particles is $A_0^{-2/3}$, while the
typical occupation number is $\lambda^{-1}A_0^{14/3}$.  The density
of the condensate is $\lambda^{-1}A_0^{8/3}$.

\subsection{Final step of thermalization: $A_0^{-8}\ll t\ll
A_0^{1/6}\lambda^{-7/4}$}

At $t\sim A_0^{-8}$, the flux of particles coming to the condensate
from the hard modes becomes comparable to the lost of particles due to
the parametric resonance.  One can expect, therefore, that the decay
of $n_0$ is now slower than $t^{-1/3}$.  We make the following
assumptions in the regime $t\gg A_0^{-8}$:

1. $n_0\sim t^{-r}$, where $r$ is some unknown constant.  

2. The particle distribution outside the condensate has the
self--similar form (\ref{self_sim}) with some $p$ and $q$.

3. The loss of particles by condensate evaporation is compensated
almost exactly by Bose condensation.  The change of $n_0$ with time is
due to a small discrepancy between the rates of the two processes.

4. The four--particle collisions can be neglected compared to
three--particle ones.

Let us find $p$, $q$ and $r$.  First, since the four--particle
scattering can be neglected, the Boltzmann equation has the same form
as eq.(\ref{Boltz_1/3}), where the factor of $t^{-1/3}$ in the r.h.s.\
is replaced by $t^{-r}$, so
\begin{equation}
  p+q+r=1
  \label{p+q+r=1}
\end{equation}
As before, the energy conservation implies
\begin{equation}
  q=4p
  \label{q=4p}
\end{equation}
Finally, the rate of particle creation by parametric resonance is
$n_0^4\sim t^{-4r}$.  On other hand, the number of particles outside
the condensate is $t^{-p}$, so the flux of non--condensate particles
coming back to the condensate is $t^{-(1+p)}$.  According to our third
assumption, the following relation should hold.
\begin{equation}
  4r=1+p
  \label{4r=1+p}
\end{equation}
From eqs.(\ref{p+q+r=1}), (\ref{q=4p}) and (\ref{4r=1+p}), one finds
$p=1/7$, $q=4/7$, $r=2/7$.  Matching with the previous regime at
$t\sim A_0^{-8}$, the following formula is obtained
\[
  n_{\bf k}(t)={1\over\lambda}\cdot{A_0^{2/21}\over t^{4/7}}
  f\left({k\over A_0^{10/21}t^{1/7}}\right)
\]
\[
  n_0\sim{1\over\lambda}\cdot{A_0^{8/21}\over t^{2/7}}
\]
As anticipated, $n_0$ decays slower than $t^{-1/3}$.  Now one can
verify our starting assumptions.  First,
$\dot{n}_0\sim\lambda^{-1}A_0^{8/21}/t^{9/7}$ is much smaller than the
rate of particle creation by parametric resonance
$n_0^4\sim\lambda^{-1}A_0^{32/21}/t^{8/7}$ if $t\gg A_0^{-8}$.
Therefore, particle loss is much smaller than the rate of evaporation
by parametric resonance, which implies that the latter is compensated
almost exactly by Bose condensation.  That the four--particle
collisions can be neglected compared with three--particle ones can be
checked easily: the l.h.s.\ of eq.(\ref{4part}) is
$A_0^{8/21}/t^{2/7}$, while the r.h.s.\ is $A_0^{22/21}/t^{2/7}$.
Since $A_0\ll1$, this condition is always satisfied.  Therefore, all
our starting assumptions are justified a posteriori.  It is
interesting to note that the four--particle scattering is negligible
in the whole history of the evolution of the system: even when the
density of the condensate is small, the dominant relaxation processes
are those involving one condensate particle.  The remnant of
homogeneous oscillations, so, play an important role in the
thermalization.

The thermal equilibrium is achieved when the typical occupation number
is of order 1.  This occurs at the the time moment
\[
  t\sim{A_0^{1/6}\over\lambda^{7/4}}
\]
Naturally, the smaller the coupling, the larger the time required for
thermalization.  The residual particle density in the condensate at
these times is
\[
  n_0\sim {A_0^{1/3}\over\lambda^{1/2}}
\]
For comparison, the density of particles outside the condensate is
$A_0^{3/2}\lambda^{-3/4}$, which is much larger than $n_0$.  Naturally,
the condensate density will decrease further and tend to 0.

\section{Conclusion}

In this paper we have followed the evolution of the scalar field from
the initial homogeneous oscillations to thermalization.  We found that
this evolutions is a rather complex process, occurring through
following regimes:

1. $t<t_0\sim A_0^{-4}$: exponential enhancement of resonance modes.
This is the feature that makes our model similar to the ones
considered in Ref.\cite{KLS}.

2. $A_0^{-4}\ll t\ll A_0^{-5}$: particles with energy
$\omega_k\approx2$ are produced with a roughly constant rate.

3. $A_0^{-5}\ll t\ll A_0^{-6}$: the typical energy of non--condensate
particles is $k_0\sim A_0^2t^{2/5}$, and the typical occupation number
is $n_{\bf k}\sim\lambda^{-1}t^{-3/5}$.  Most energy is still
concentrated in the condensate, whose density remains approximately
constant, $n_0\sim\lambda^{-1}A_0^2$.

4. $A_0^{-6}\ll t\ll A_0^{-8}$: the condensate density decays as
$n_0\sim\lambda^{-1}t^{-1/3}$.  The typical energy and occupation
number of non--condensate particles are $k_0\sim A_0^{2/5}t^{2/15}$
and $n_{\bf k}\sim\lambda^{-1}A_0^{2/5}t^{-8/5}$, respectively.

5. $A^{-8}\ll t\ll A^{1/6}\lambda^{-7/4}$:
$n_0\sim\lambda^{-1}A_0^{8/21}t^{-2/7}$, $k_0\sim A_0^{10/21}t^{1/7}$,
$n_{\bf k}\sim\lambda^{-1}A_0^{2/21}t^{-4/7}$,

The thermalization finishes at $t\sim A^{1/6}\lambda^{-7/4}$.

We see that in our model the Hartree--Fock approximation works only in
the regime 1 and 2, $t\ll A_0^{-5}$.  At $A_0^{-5}\ll t\ll A_0^{-8}$
our calculations give the rate of particle loss from the condensate
twice smaller than predicted by the Hartree--Fock approximation.
Finally, at $t\gg A_0^{-8}$ the scattering changes the decay law of
the condensate, so instead of $n_0\sim t^{-1/3}$ one has $n_0\sim
t^{-2/7}$.

Though all calculations in this paper are based on the existence of
the additional small parameter $A_0$, one may expect that some
features similar to those discovered in our model also occur in the
some of the realistic theories.  As one of them, we speculate that the
exponential enhancement of the parametric resonance modes is
terminated by the back--reaction of produced modes and the coherent
nature of the latter may slow down the energy loss of the homogeneous
oscillations.  It is also natural to argue that the thermalization
process takes a long time, during which some types of self--similar
behavior of the distribution functions may have place in some models.
Hopefully, the techniques used in this paper may prove helpful in the
understanding of the dynamics of the inflaton in these theories, which
would provide more accurate physical predictions than those provided
by the Hartree--Fock approximation.

It is particularly interesting to compare our analytical results with
the numerical simulations in the massless $\phi^4$ theory performed in
Ref.\cite{Khlebnikov}.  It is worth noting that though the massless
$\phi^4$ theory lacks any additional small parameter beside $\lambda$,
the width of the resonance band, as well as the value of the maximal
index of exponential enhancement $s_k$ are numerically small in this
model.  Therefore, one can expect that many qualitative features found
in our model are also valid for the massless $\phi^4$ theory.  Though
late thermalization epochs are not considered in
Ref.\cite{Khlebnikov}, the behavior of the field show many similarity
with our model in earlier regimes.  In particular, the fluctuations of
the field comming from the non--condensate modes are relatively small
at the end of the linear regime.  The peaks in the spectrum are also
found to be present during some time interval and they are shown to be
caused by scattering processes.  From the analogy with our model, the
bounces found in the time evolution of the amplitude of homogeneous
oscillations can be attributed to the coherence of resonance modes,
not to Bose condensation as argued in Ref.\cite{Khlebnikov}.

The author thanks P.~Arnold and L.~Yaffe for stimulating discussions,
and L.~Kofman and A.~Linde for explaining many points of
Ref.\cite{KLS}.  I thank S.Yu.~Khlebnikov and I.I.~Tkachev for
numerous comments on the manuscript.  This works is supported, in
part, by the U.S. Department of Energy under grant
\#DE--FG03--96ER40956.

\appendix

\section{Derivation of Hartree--Fock equations for 
\lowercase{$\alpha_i(k)$}}

In this appendix we will show how eq.(\ref{HF_main}) can be derived.
For convenience, we will to work with complex notations.  In analogy
with the technique used in Sect.2 to find the first and second
resonance band, we will try to find the solution to the Hartree--Fock
equation in the following form,
\[
  \phi_0={1\over\sqrt{\lambda}}\left(
  A\e^{-i\Omega}+B\e^{-3i\Omega}+\mbox{h.c.}+\cdots\right)
\]
\begin{equation}
  F_i(k)={1\over\sqrt{\lambda}}\left(
  \alpha_i(k)\e^{-2i\Omega}+\beta_i(k)\e^{-4i\Omega}
  +\delta_i(k)\e^{-6i\Omega}+\mbox{h.c.}+\gamma_i(k)+\cdots\right)
  \label{app_ansatz}
\end{equation}
where
\[
  \Omega=\int\!dt\,\omega(t)
\]
and $\omega(t)$ is the time--dependent frequency of the oscillations
that is chosen so that $A$ is real.  The order of magnitude of the
parameters in eq.(\ref{app_ansatz}) are expected to be as follows (the
consistence of these relations can be verified a posteriori),
\[
  B\sim O(A^3),\qquad \dot{A}\sim O(A^7),
\]
\[
  \alpha_i(k)\sim O(A^2), \qquad \beta_i(k)\sim \gamma_i(k)\sim O(A^4),
  \qquad \delta_i(k)\sim O(A^6)
\]
\[
  \dot{\alpha}_i(k)\sim O(A^6)
\]
\[
  \omega-1\sim O(A^2),\qquad \dot{\omega}\sim O(A^8)
\]

Let us substitute the ansatz (\ref{app_ansatz}) to the Hartree--Fock
equations, eqs.(\ref{HF_eq1}) and (\ref{HF_eq2}).  First let us
consider eq.(\ref{HF_eq2}).  To the order of $A^6$, the l.h.s.\ is
expanded into terms proportional to $1$, $\e^{\pm 2i\Omega}$, $\e^{\pm
4i\Omega}$ and $\e^{\pm 6i\Omega}$.  Each coefficient of these
exponents must be equal 0, which gives us 7 equations.  Those
corresponding to $\e^{\pm 6i\Omega}$ are the equations that determine
$d_{\bf k}$.  Since $d_{\bf k}$ do not emerge in other equations,
these equations can be ignored.  It turns out that we will need the
coefficient of 1 and $\e^{\pm4i\Omega}$ only with the accuracy of
$O(A^4)$ (not $O(A^6)$).  To this order, terms proportional to 1 read
\[
  4\gamma_i(k)+{A^2\over2}(\alpha_i(k)+\alpha^*_i(k))=0
\]
while the equations corresponding to $\e^{\pm4i\Omega}$ are
\[
  -12\beta_i(k)+{A^2\over2}\alpha_i(k)=0
\]
and its complex conjugation.  These equations relate $\beta_i(k)$ and
$\gamma_i(k)$ with $\alpha_i(k)$,
\[
  \beta_i(k)={A^2\over24}\alpha_i(k)
\]
\begin{equation}
  \gamma_i(k)=-{A^2\over8}(\alpha_i(k)+\alpha^*_i(k))
  \label{app_bc}
\end{equation}
The term $\e^{-2i\Omega}$ implies the following equation
\begin{equation}
  -4i\dot{\alpha}_i(k)+
  \left(-4\omega^2+\omega_k^2+A^2+{\cal I}\right)\alpha_i(k)+
  \left(AB+{\cal C}\right)\alpha^*_i(k)+
  {A^2\over2}(\beta_i(k)+\gamma_i(k))=0
  \label{app*}
\end{equation}
where we introduce the notations
\[
  {\cal I}=\int\!{d{\bf q}\over(2\pi)^32\omega_q}\,
  (|\alpha_1(q)|^2+|\alpha_2(q)|^2)
\]
\[
  {\cal C}={1\over2}\int\!{d{\bf q}\over(2\pi)^32\omega_q}\,
  (\alpha^2_1(q)+\alpha^2_2(q))
\]
Eq.(\ref{app*}) can be rewritten in the following form after taking
into account eqs.(\ref{app_bc})
\begin{equation}
  -4i\dot{\alpha}_i(k)+\left(-4\omega^2+\omega_k^2+A^2-{A^4\over24}+
  {\cal I}\right)\alpha_i(k)+
  \left(AB-{A^4\over16}+{\cal C}\right)\alpha^*_i(k)=0
  \label{app_a1}
\end{equation}

Now let us turn to eq.(\ref{HF_eq1}).  To the order of $O(A^7)$, both
sides can be expanded in terms proportional to $\e^{\pm i\Omega}$,
$\e^{\pm3i\Omega}$, $\e^{\pm5i\Omega}$ and $\e^{\pm7i\Omega}$.  We
will not need the equation corresponding to $\e^{\pm5i\Omega}$ and
$\e^{\pm7i\Omega}$.  Moreover we will need the equation coming from
the term $\e^{-3i\Omega}$ with the accuracy of $O(A^5)$ only, which
has the form,
\[
  (-9\omega^2+1+A^2)B+{A^3\over6}+A{\cal C}=0
\]
form which one obtains,
\begin{equation}
  B={A^3\over6(9\omega^2-1-A^2)}+{A\over8}{\cal C}
  \label{app_B}
\end{equation}
The equation coming from the terms containing $\e^{-i\Omega}$ is,
\[
  -2i\dot{A}-\omega^2A+A+{A^3\over2}+{A^2B\over2}+AB^*B+A{\cal I}+
  B^*{\cal C}+
\]
\[
  A\int{d{\bf q}\over(2\pi)^32\omega_q}
  (\alpha^*_i(q)\beta_i(q)+\alpha_i(q)\gamma_i(q))=0
\]
Together with eqs.(\ref{app_bc}), (\ref{app_B}), and the condition
that $A$ is real, one obtain the formula for $\omega^2$,
\begin{equation}
  \omega^2=1+{A^2\over2}+{A^4\over96}+{\cal I}+O(A^6)
  \label{app_omega}
\end{equation}
and the equation that $A$ satisfies,
\begin{equation}
  i\dot{A}+{A^3\over24}({\cal C}-{\cal C}^*)=0
  \label{app_A}
\end{equation}
Taking into account eqs.(\ref{app_B}) and (\ref{app_omega}),
eq.(\ref{app_a1}) can be rewritten as,
\begin{equation}
  -4i\dot{\alpha}_i(k)+(\omega_k^2-4-A^2-{A^4\over12}-3{\cal I})
  \alpha_i(k)-{A^4\over24}\alpha^*_i(k)+{\cal C}\alpha^*_i(k)=0
  \label{app_a2}
\end{equation}
From eq.(\ref{app_A}) and eq.(\ref{app_a2}), it is easy to verify the
energy conservation,
\[
  {d\over dt}(A^2+4{\cal I})=0
\]
So, one has
\[
  A^2+4{\cal I}=A_0^2
\]
where $A_0$ is the value of $A$ at $t=0$ when $a_{\bf k}$ are still 
very small.  Eq.(\ref{app_a2}) now can be rewritten in
the final form,
\[
  -4i\dot{\alpha}_i(k)+(\omega^2-\omega_c^2+{\cal I})\alpha_i(k)-
  {A_0^4\over24}\alpha^*_i(k)+{\cal C}\alpha^*_i(k)=0
\]
where
\[
  \omega^2_c=4+A_0^2+{1\over12}A_0^4
\]
is the center of the second resonance band in the linear regime.

\section{Derivation of Boltzmann equation for \lowercase{$n_l$}}

Let us substitute eq.(\ref{Boltz_ansatz}) to eq.(\ref{nnn}).  First
consider the contribution from the first term in the parenthesis in
eq.(\ref{nnn}),
\[
  \int\!{d{\bf k}_1d{\bf k}_2d{\bf k}_3\over
  \omega_k\omega_{k_1}\omega_{k_2}\omega_{k_3}}\,
  n_{{\bf k}_1}n_{{\bf k}_2}n_{{\bf k}_3}
  \delta({\bf k}_1+{\bf k}_2-{\bf k}_3{\bf k})
  \delta(\omega_{k_1}+\omega_{k_2}-\omega_{k_3}-\omega_k)
\]
The main contribution to this integral correspond to the case when one
of the momenta ${\bf k}_1$, ${\bf k}_2$, or ${\bf k}_3$ are equal 0.
The two former cases yields the same result, so this integral is
reduced to
\[
  {n_0\over\omega_k}\left(\int\!
  {d{\bf k}_1d{\bf k}_2\over\omega_{k_1}\omega_{k_2}}\,
  n_{k_1}n_{k_2}\delta({\bf k}_1+{\bf k}_2-{\bf k})
  \delta({\bf k}_1+{\bf k}_2-{\bf k})
  \delta(\omega_{k_1}+\omega_{k_2}-\omega_k-1)\right.+
\]
\begin{equation}
  +\left.2\int\!{d{\bf k}_1d{\bf k}_3\over\omega_{k_1}\omega_{k_3}}\,
  n_{k_1}n_{k_3}\delta({\bf k}_1-{\bf k}_3-{\bf k})
  \delta({\bf k}_1-{\bf k}_3-{\bf k})
  \delta(\omega_{k_1}+1-\omega_{k_3}-\omega_k)\right)
  \label{B:1st_contr}
\end{equation}
Let us evaluate the first integral in eq.(\ref{B:1st_contr}).  The
integral over ${\bf k}_2$ can be taken easily due to the delta
function $\delta({\bf k}_1+{\bf k}_2-{\bf k})$.  Recalling
eq.(\ref{Boltz_ansatz}), this integral can be rewritten into the form,
\[
  \sum_{l_1,l_2}{n_{l_1}n_{l_2}\over(4\pi)^2k^2_{l_1}k^2_{l_2}}
  (2\pi)\int\!k_1^2dk_1\sin\,\theta d\theta\,\delta(k_1-k_{l_1})
  \delta(|{\bf k}-{\bf k}_1|-k_{l_2})\delta(l_1+l_2-1-\omega_k)
\]
where $\theta$ is the angle between ${\bf k}$ and ${\bf k}_1$ so
$|{\bf k}-{\bf k}_1|=\sqrt{k^2+k_1^2-2kk_1\cos\theta}$.  Now the
integral over $dk_1$ and $d\theta$ can be taken, and the result reads
\[
  {n_0\over8\pi k^2}\sum_{l_1,l_2}{n_{l_1}n_{l_2}\over k_{l_1}k_{l_2}
  l_1l_2}\delta\left(k-\sqrt{(l_1+l_2-1)^2-1}\right)
  ={n_0\over8\pi k^2}\sum_l\delta(k-k_l)
   \sum_{l_1+l_2=l+1}{n_{l_1}n_{l_2}\over k_{l_1}k_{l_2}l_1l_2}
\]
The second contribution in eq.(\ref{B:1st_contr}) can be evaluated in
a similar way.  Therefore one obtains
\[
  \int\!{d{\bf k}_1d{\bf k}_2d{\bf k}_3\over
  \omega_k\omega_{k_1}\omega_{k_2}\omega_{k_3}}\,
  n_{{\bf k}_1}n_{{\bf k}_2}n_{{\bf k}_3}
  \delta({\bf k}_1+{\bf k}_2-{\bf k}_3-{\bf k})
  \delta(\omega_{k_1}+\omega_{k_2}-\omega_{k_3}-\omega_k)=
\]
\[
  ={n_0\over4\pi k^2}\sum_l\delta(k-k_l)\left(
  {1\over2}\sum_{l_1+l_2=l+1}{n_{l_1}n_{l_2}\over k_{l_1}k_{l_2}l_1l_2}
  +\sum_{l_1-l_2=l_1}{n_{l_1}n_{l_2}\over k_{l_1}k_{l_2}l_1l_2}\right)
\]
Using the same technique, the contribution from the term $n_{{\bf
k}_1}n_{{\bf k}_2}n_{\bf k}$ in eq.(\ref{nnn}) is
\[
  \int\!{d{\bf k}_1d{\bf k}_2d{\bf k}_3\over
  \omega_k\omega_{k_1}\omega_{k_2}\omega_{k_3}}\,
  n_{{\bf k}_1}n_{{\bf k}_2}n_{\bf k}
  \delta({\bf k}_1+{\bf k}_2-{\bf k}_3-{\bf k})
  \delta(\omega_{k_1}+\omega_{k_2}-\omega_{k_3}-\omega_k)
  ={n_0n_{\bf k}\over k\omega_k}\sum_{l_1>\omega_k}
  {n_{l_1}\over k_{l_1}l_1}
\]
Analogously, the last two terms in eq.(\ref{nnn}) give
\[
  -{n_0n_{\bf k}\over k\omega_k}\left(\sum_{l_1=2}^\infty
  {n_{l_1}\over k_{l_1}l_1}+
  \sum_{l_1<\omega_k}{n_{l_1}\over k_{l_1}l_1}\right)
\]
Collecting all the contributions and recalling
eq.(\ref{Boltz_ansatz}), one finds the Boltzmann equation for $n_l$,
\[
  {dn_l\over dt}=\lambda^2n_0\left(\sum_{l_1+l_2=l+1}
  {n_{l_1}n_{l_2}\over k_{l_1}k_{l_2}l_1l_2}+
  \sum_{l_1-l_2=l-1}{n_{l_1}n_{l_2}\over k_{l_1}k_{l_2}l_1l_2}
  -{n_l^2\over k_l^2l^2}-{2n_l\over k_ll}
  \sum_{l_1<l}{n_l\over k_ll}\right)+j\delta_{l2}
\]
where the last term in the r.h.s.\ comes from the Hartree--Fock
particle production.

\newpage

\begin{figure}
\leavevmode
\epsffile{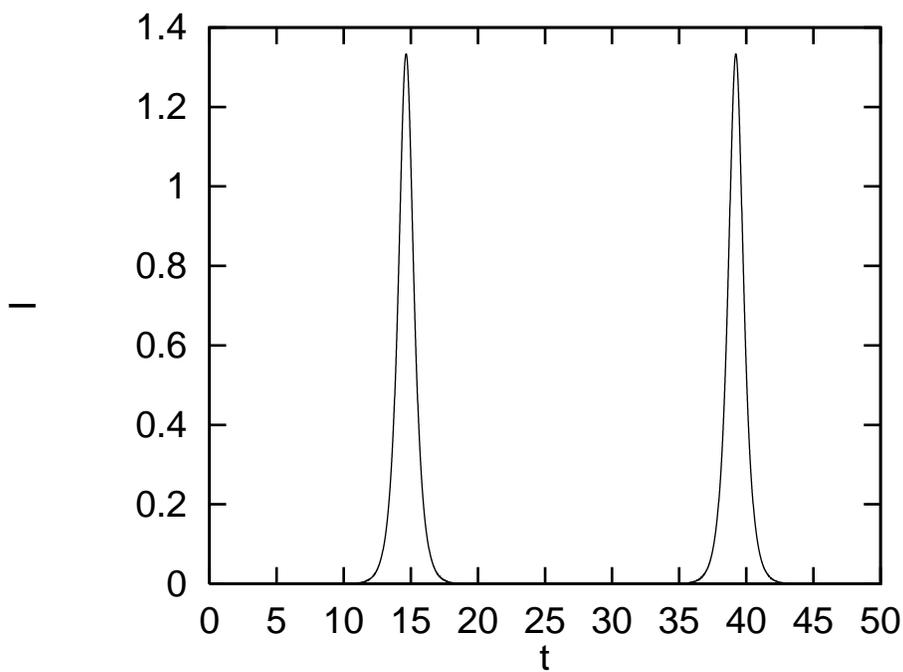}
\vspace{0.5cm}
\caption{The dependence of $I$ on $t$ in the limit
$\log(1/\lambda)\to\infty$.  The regime of validity of the
approximation contains only the first peak.}
\end{figure}

\begin{figure}
\leavevmode
\epsffile{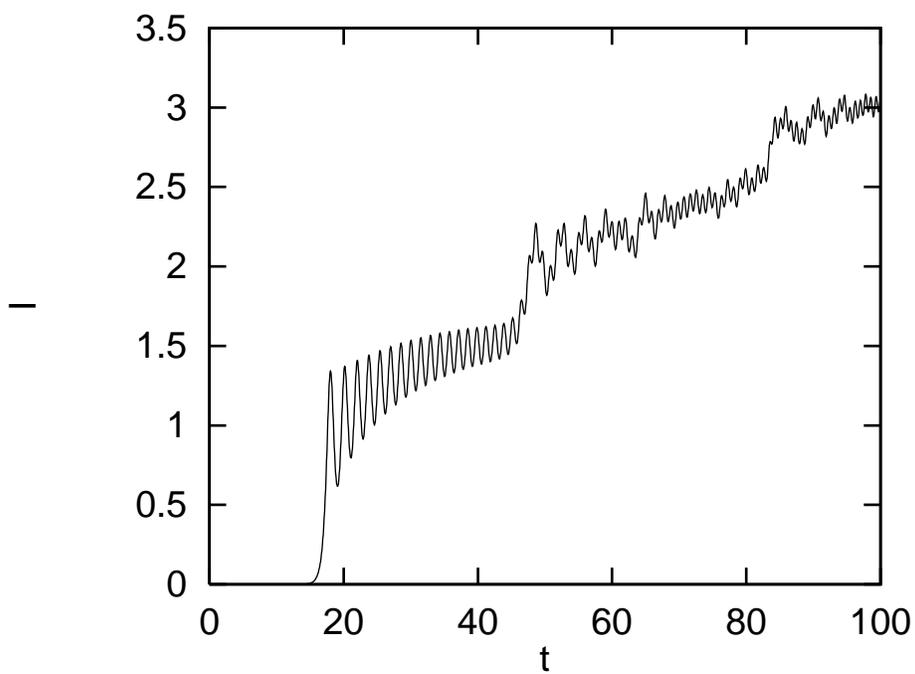}
\vspace{0.5cm}
\caption{The dependence of total energy of inhomogeneous modes on time
in the Hartree--Fock approximation.}
\end{figure}

\end{document}